\documentclass[iop]{emulateapj} 
\usepackage{epstopdf}
\usepackage{epsfig} 
\usepackage{morefloats}
\usepackage[backref,breaklinks,colorlinks,citecolor=blue]{hyperref}
\usepackage[all]{hypcap}

\def\msun{\mbox{$M_\odot$}} 
\def\lsun{\mbox{$L_\odot$}}
\def\rsun{\mbox{$R_\odot$}} 

\def\teff{\mbox{$T_{\rm eff}$}} 
\def\logg{\mbox{$\log g$}}
\def\fe{\mbox{[Fe/H]}} 
\def\feh{\mbox{[Fe/H]}}

\def\ms{\mbox{m s$^{-1}$}}

\def\mstar{M$_{\star}$} 
\def\lstar{L$_{\star}$} 
\def\rstar{R$_{\star}$} 
\def\mjup{$M_{\rm Jup}$}

\def\vsini{$v \sin i$} 
\def\msini{$M_{P} \sin{i}$}

\def\hipp{$Hipparcos$} 
\def\snr{\mbox{\rm signal-to-noise ratio}}
\def\caii{\ion{Ca}{2}} 
\def\shk{\mbox{$S_{\rm HK}$}}
\def\chinu{$\chi_{\nu}^{2}$} 
 
\def\sme{\textit{SME}}
\def\yy{\mbox{$Y^{2}$}}
\def\rms{rms} 
\def\kfme{\textit{KFME}}

\def\spectypefive{K3} \def\lumclassfive{subgiant} 
\def\vmagfive{$8.05$}
\def\absmagfive{$2.76$} 
\def\bvfive{$0.99$} 
\def\bcfive{$-0.30$}
\def\distfive{$114$} 
\def\distuncfive{$11$} 
\def\tefffive{$4958$} \def\teffuncfive{$44$} 
\def\loggfive{$3.45$} \def\logguncfive{$0.06$}
\def\fehfive{$0.15$} \def\fehuncfive{$0.03$}
\def\mstfive{$1.51$} \def\mstuncfive{$0.11$} 
\def\rstfive{$3.85$} \def\rstuncfive{$0.40$} 
\def\lstfive{$8.1$} \def\lstuncfive{$1.6$}
\def\agefive{$3.1$} \def\ageuncfive{$0.6$} 
\def\rhkfive{$-5.28$}

\def\nobsvho{$81$} \def\jittervho{$4.2$}
\def\rmsvho{$7.18$} \def\chinuvho{$2.64$}
 
\def\orbperbvho{$641$} \def\orbperuncbvho{$2$}
\def\orbtpbvho{$16288$} \def\orbtpuncbvho{$790$}
\def\orbebvho{$0.02$} \def\orbeuncbvho{$0.03$}
\def\orbombvho{$97$} \def\orbomuncbvho{$90$}
\def\orbmbvho{$1.76$} \def\orbmuncbvho{$0.07$}
\def\orbaplbvho{$1.6697$} \def\orbapluncbvho{$0.0036$}
\def\orbkbvho{$31.6$} \def\orbkuncbvho{$1.2$}

\def\orbpercvho{$886$} \def\orbperunccvho{$8$}
\def\orbtpcvho{$13453$} \def\orbtpunccvho{$92$}
\def\orbecvho{$0.15$} \def\orbeunccvho{$0.06$}
\def\orbomcvho{$252$} \def\orbomunccvho{$34$}
\def\orbmcvho{$1.15$} \def\orbmunccvho{$0.08$}
\def\orbaplcvho{$2.071$} \def\orbaplunccvho{$0.013$}
\def\orbkcvho{$18.8$} \def\orbkunccvho{$1.3$}

\def\vmageleven{$7.51$} 
\def\absmageleven{$3.94$} 
\def\bveleven{$0.61$}  
\def\bceleven{$-0.04$} 
\def\disteleven{$51.7$} \def\distunceleven{$1.6$} 
\def\teffeleven{$6055$} \def\teffunceleven{$44$} 
\def\loggeleven{$4.30$} \def\loggunceleven{$0.06$} 
\def\feheleven{$0.29$} \def\fehunceleven{$0.03$}
\def\msteleven{$1.24$} \def\mstunceleven{$0.02$} 
\def\rsteleven{$1.33$} \def\rstunceleven{$0.05$} 
\def\lsteleven{$2.13$} \def\lstunceleven{$0.16$} 
\def\ageeleven{$2.4$} \def\ageunceleven{$0.6$}
\def\rhkeleven{$-4.98$} 

\def\dvdtoov{$-7.19$} \def\dvdtuncoov{$0.26$}
\def\nobsoov{$107$} \def\jitteroov{$2.8$}
\def\rmsoov{$4.80$} \def\chinuoov{$2.20$}
 
\def\orbperboov{$1627.5$} \def\orbperuncboov{$5.9$}
\def\orbtpboov{$16637.2$} \def\orbtpuncboov{$7.6$}
\def\orbeboov{$0.37$} \def\orbeuncboov{$0.01$}
\def\orbomboov{$218.9$} \def\orbomuncboov{$1.6$}
\def\orbmboov{$4.21$} \def\orbmuncboov{$0.07$}
\def\orbaplboov{$2.708$} \def\orbapluncboov{$0.007$}
\def\orbkboov{$78.4$} \def\orbkuncboov{$1.2$}

\def\orbpercoov{$223.6$} \def\orbperunccoov{$0.6$}
\def\orbtpcoov{$14127$} \def\orbtpunccoov{$9$}
\def\orbecoov{$0.24$} \def\orbeunccoov{$0.05$}
\def\orbomcoov{$272$} \def\orbomunccoov{$15$}
\def\orbmcoov{$0.36$} \def\orbmunccoov{$0.02$}
\def\orbaplcoov{$0.721$} \def\orbaplunccoov{$0.001$}
\def\orbkcoov{$12.5$} \def\orbkunccoov{$0.7$}

\def\vmagseven{$7.84$} 
\def\absmagseven{$3.15$}
 \def\bvseven{$0.99$}
\def\bcseven{$-0.32$} 
\def\parseven{$11.54$} \def\paruncseven{$0.83$}
\def\distseven{$68.7$} \def\distuncseven{$6.2$}
\def\teffseven{$4917$} \def\teffuncseven{$44$} 
\def\loggseven{$3.56$} \def\logguncseven{$0.06$} 
\def\fehseven{$0.25$} \def\fehuncseven{$0.03$}
\def\mstseven{$1.41$} \def\mstuncseven{$0.08$} 
\def\rstseven{$3.3$} \def\rstuncseven{$0.3$} 
\def\lstseven{$5.7$} \def\lstuncseven{$0.9$}
\def\ageseven{$4.0$} \def\ageuncseven{$0.7$} 
\def\rhkseven{$-5.24$}

\def\spectypeseven{K3}

\def\nobssvs{$41$} \def\jittersvs{$4.3$}
\def\rmssvs{$4.63$} \def\chinusvs{$1.09$}
 
\def\orbperbsvs{$5040$} \def\orbperuncbsvs{$3414$}
\def\orbtpbsvs{$19655$} \def\orbtpuncbsvs{$2297$}
\def\orbebsvs{$0.36$} \def\orbeuncbsvs{$0.16$}
\def\orbombsvs{$301$} \def\orbomuncbsvs{$75$}
\def\orbmbsvs{$5.6$} \def\orbmuncbsvs{$1.2$}
\def\orbaplbsvs{$6.5$} \def\orbapluncbsvs{$2.0$}
\def\orbkbsvs{$57$} \def\orbkuncbsvs{$11$}

\def\orbpercsvs{$341.7$} \def\orbperunccsvs{$6.1$}
\def\orbtpcsvs{$14411$} \def\orbtpunccsvs{$172$}
\def\orbecsvs{$0.13$} \def\orbeunccsvs{$0.10$}
\def\orbomcsvs{$29$} \def\orbomunccsvs{$136$}
\def\orbmcsvs{$1.15$} \def\orbmunccsvs{$0.30$}
\def\orbaplcsvs{$1.073$} \def\orbaplunccsvs{$0.013$}
\def\orbkcsvs{$26.7$} \def\orbkunccsvs{$6.6$}

\def\vmagten{$7.84$}  
\def\absmagten{$$}  
\def\bvten{$0.93$}
\def\bcten{$-0.27$} 
\def\distten{$$} \def\distuncten{$$}
\def\tefften{$5034$} \def\teffuncten{$44$} 
\def\loggten{$3.50$} \def\logguncten{$0.06$} 
\def\fehten{$0.11$} \def\fehuncten{$0.03$}
\def\mstten{$1.56$} \def\mstuncten{$0.09$} 
\def\rstten{$$}\def\rstuncten{$$} 
\def\lstten{$$} \def\lstuncten{$$}
\def\ageten{$$} \def\ageuncten{$$} 
\def\rhkten{$-5.17$}
 
\def\spectypeten{K2}

\def\nobsozu{$43$} \def\jitterozu{$4.7$}
\def\rmsozu{$5.98$} \def\chinuozu{$1.53$}
 
\def\orbperbozu{$1043$} \def\orbperuncbozu{$9$}
\def\orbtpbozu{$17062$} \def\orbtpuncbozu{$770$}
\def\orbebozu{$0.11$} \def\orbeuncbozu{$0.06$}
\def\orbombozu{$198$} \def\orbomuncbozu{$60$}
\def\orbmbozu{$2.10$} \def\orbmuncbozu{$0.15$}
\def\orbaplbozu{$2.335$} \def\orbapluncbozu{$0.014$}
\def\orbkbozu{$31.5$} \def\orbkuncbozu{$2.2$}

\shorttitle{Four New Planets from N2K} \shortauthors{Giguere et al.} 

\begin{document}

\title{Newly-Discovered Planets Orbiting HD~5319, HD~11506,  HD~75784 and HD~10442 from the N2K Consortium\altaffilmark{*}}

\author{Matthew J. Giguere\altaffilmark{1}, Debra A.
Fischer\altaffilmark{1}, 
Matthew J. Payne\altaffilmark{2}, 
John M. Brewer\altaffilmark{1}, 
John Asher Johnson\altaffilmark{2}, 
Andrew W. Howard\altaffilmark{3}, 
Howard T. Isaacson\altaffilmark{4}
}

\begin{abstract} 
Initially designed to discover short-period planets, the N2K campaign has since evolved to discover new worlds at large separations from their host stars. Detecting such worlds will help determine the giant planet occurrence at semi-major axes beyond the ice line, where gas giants are thought to mostly form. Here we report four newly-discovered gas giant planets (with minimum masses ranging from 0.4 to 2.1 \mjup) orbiting stars monitored as part of the N2K program. Two of these planets orbit stars already known to host planets: HD~5319 and HD~11506. The remaining discoveries reside in previously-unknown planetary systems: HD~10442 and HD~75784. The refined orbital period of the inner planet orbiting HD~5319 is 641 days. The newly-discovered outer planet orbits in \orbpercvho \ days. The large masses combined with the proximity to a 4:3 mean motion resonance make this system a challenge to explain with current formation and migration theories. HD~11506 has one confirmed planet, and here we confirm a second. The outer planet has an orbital period of \orbperboov \ days, and the newly-discovered inner planet orbits in \orbpercoov \ days. A planet has also been discovered orbiting HD~75784 with an orbital period of \orbpercsvs \ days. There is evidence for a longer period signal; however, several more years of observations are needed to put tight constraints on the Keplerian parameters for the outer planet. Lastly, an additional planet has been detected orbiting HD~10442 with a period of \orbperbozu \ days.
\end{abstract}


\altaffiltext{*}{Based on observations obtained at the W. M. Keck
Observatory, which is operated by the University of California and the
California Institute of Technology. Keck time has been granted by NOAO
and NASA.}

\altaffiltext{1}{Department of Astronomy, Yale University, 260 Whitney Ave, New Haven, CT 06511, USA}
\altaffiltext{2}{Harvard-Smithsonian Center for Astrophysics, 60 Garden Street, Cambridge, MA 02138, USA}
\altaffiltext{3}{Institute for Astronomy, University of Hawaii at Manoa, Honolulu, HI, USA}
\altaffiltext{4}{Department of Astronomy, University of California, Berkeley, Berkeley, California 94720, USA}

\section{Introduction}

	Many details concerning planet formation and evolution have been gleaned from the ensemble of observed extrasolar planetary systems. An early example is the paradigm shift caused by the first few systems discovered, which contained gas giant planets orbiting well within the snow line \citep{1995Natur.378..355M, 1996ApJ...464L.147M}. This challenged the prevailing planet formation theory, in which planets form and remain several astronomical units (AU) from their parent stars \citep{1993ARA&A..31..129L, 1996Icar..124...62P}. Planet formation theory quickly evolved to explain the newly-discovered systems in terms of migration \citep{1996Natur.380..606L}. 

	As the number of discovered planetary systems accumulated, new connections between stellar parameters and the occurrence of planets became apparent. With only seven extrasolar planetary systems known at the time, \citet{1997MNRAS.285..403G} noted a shared characteristic that four of the host stars had super-solar metallicities. As the number of known planetary systems grew, the connection between host star metallicity and giant planet occurrence became more pronounced. This culminated with \citet{2004A&A...415.1153S} measuring elemental abundances of 139 stars (98 known to host giant planets and 41 with no known companions), and \citet{2005ApJ...622.1102F} performing a thorough statistical analysis of 850 FGK-type stars, revealing the giant planet-metallicity correlation. 

	The more favorable detection rate for gas giant planets orbiting metal-rich stars combined with the exceptional scientific pay-off of transiting planets inspired a focused search for short-period gas giant planets orbiting metal-rich stars: the N2K (Next 2000 target stars) Doppler Survey \citep{2005ApJ...620..481F}. Some particularly interesting transiting planets discovered as part of this program include HD~17156~b, a highly-eccentric transiting planet \citep{2007ApJ...669.1336F, 2007A&A...476L..13B, 2010ApJ...719..602S, 2011ASPC..450...71L}, and HD~149026~b, a surprisingly dense transiting hot-Jupiter \citep{2005ApJ...633..465S, 2006ApJ...642..495F, 2009ApJ...695L.159D, 2012PASP..124..809A}. Doppler measurements of these systems can reveal valuable information concerning their migration history. 
	
\capstartfalse
\begin{deluxetable*}{l l l r}[ht]
\tablecaption{N2K Discoveries \label{tab:n2k}}
\tablewidth{0pt} 
\tablehead{ \colhead{Star ID}   & \colhead{Period}        
            & \colhead{\msini}  & \colhead{Reference} \\ 
            \colhead{ }   & \colhead{[d]}        
            & \colhead{[\mjup]}  & \colhead{ } 
            } 
\startdata
HD~86081 b   &   2.14  & 1.50 & \citet{2006ApJ...647..600J}  \\
HD~149026 b  &   2.88  & 0.36 & \citet{2005ApJ...633..465S}  \\
HD~88133 b   &   3.42  & 0.30 & \citet{2005ApJ...620..481F}  \\
HD~149143 b  &   4.07  & 1.33 & \citet{2006ApJ...637.1094F}  \\
HD~125612 c  &   4.15  & 0.06 & \citet{2007ApJ...669.1336F}  \\
HD~109749 b  &   5.24  & 0.28  & \citet{2006ApJ...637.1094F}  \\
HIP~14810 b  &   6.67  & 3.87   & \citet{2007ApJ...657..533W}  \\
HD~179079 b  &  14.5  & 0.08  & \citet{2009ApJ...702..989V}  \\
HD~33283 b   &  18.2   & 0.33  & \citet{2006ApJ...647..600J}  \\
HD~17156 b   &  21.2   & 3.30   & \citet{2007ApJ...669.1336F}  \\
HD~224693 b  &  26.7  & 0.72  & \citet{2006ApJ...647..600J}  \\
HD~163607 b  &  75.3  & 0.77  & \citet{2012ApJ...744....4G}  \\
HD~231701 b  &  142    & 1.09  & \citet{2007ApJ...669.1336F}  \\
HIP~14810 c  & 148    & 1.28  & \citet{2007ApJ...657..533W}  \\
HD~154672 b  &  164  & 5.01   & \citet{2008AJ....136.1901L}  \\
HD~11506 c   &  223  & 0.40  &   this work   \\
HD~205739 b  &  280    & 1.49   & \citet{2008AJ....136.1901L}  \\
HD~164509 b  &  282  & 0.48  & \citet{2012ApJ...744....4G}  \\
HD~75784 b   &  342    & 1.15  &  this work    \\
HD~75898 b   &  418    & 2.52   & \citet{2007ApJ...670.1391R}  \\
HD~16760 b   &  465    & 13.3 & \citet{2009ApJ...703..671S}  \\
HD~96167 b   & 499     & 0.69  & \citet{2009PASP..121..613P}  \\
HD~125612 b  &  559    & 3.07   & \citet{2007ApJ...669.1336F}  \\
HD~5319 b    &   641   & 1.76   & this work and \citet{2007ApJ...670.1391R}  \\
HD~5319 c    &  886    & 1.15  &  this work   \\
HD~16175 b   &  990    & 4.38   & \citet{2009PASP..121..613P}  \\
HD~38801 b   &  696    & 10.0  & \citet{2010ApJ...715..550H}  \\
HIP~14810 d  & 951     & 0.58   & \citet{2009ApJ...699L..97W}  \\
HD~10442 b   & 1043    & 2.10  &  this work    \\
HD~163607 c  & 1314    & 2.29  & \citet{2012ApJ...744....4G}  \\
HD~11506 b   & 1617    & 4.80   &  this work and \citet{2007ApJ...669.1336F}  \\
HD~73534 b   & 1770    & 1.07  & \citet{2009ApJ...702..989V}  \\
HD~75784 c   &  5040    & 5.6  &  this work    \\
\enddata 
\end{deluxetable*}
\capstarttrue		

	Planets are thought to migrate either through planet-disk interactions via Type I and Type II migration \citep{1980ApJ...241..425G, 1996Natur.380..606L, Ida:2004ko}, or through gravitational interactions. The latter mechanism includes interactions either with other planets in the system \citep{2008ApJ...686..621F, 2008ApJ...686..580C, 2011ApJ...735..109W}, or with other stars via the Kozai mechanism \citep{2003ApJ...589..605W, 2007ApJ...669.1298F}. If planets migrate slowly through Type I or II migration, their orbital axes remain well-aligned with the rotational axes of their host stars. If planets migrate through gravitational interactions, their orbital axes are more likely to be misaligned relative to the rotational axes of their host stars \citep{2008ApJ...686..580C}. By measuring the Rossiter-McLaughlin Effect, the spin-orbit alignment (obliquity) can be calculated for these hot Jupiter transiting systems \citep{2005ApJ...631.1215W,2010ApJ...718L.145W}, shedding light onto the dominant migration mechanism. 

	This hypothesis assumes that the protoplanetary disks from which planets are born are well-aligned with the rotational axes of the stars they surround. Based on the Solar System, this assumption appears valid \citep{1993ARA&A..31..129L}. However, this has been recently called into question in both single-star and multiple-star systems \citep{2010MNRAS.401.1505B,2011MNRAS.412.2790L, 2012Natur.491..418B}. \citet{2014MNRAS.438L..31G} have examined 11 single-star systems with \textit{Herschel} where the stellar inclination was known and the surrounding dust belts were spatially resolved. They found that all 11 disk-star spin angles were well-aligned, showing that misalignment mechanisms operate rarely in single star systems. Additionally, there are currently at least two observational programs exploring misalignment in binary systems: one is looking at the spin-orbit alignments in eclipsing binary star systems \citep{2013ApJ...767...32A}, while the other is searching for unbeknownst widely-separated massive companions in transiting hot Jupiter systems where obliquities have been measured \citep{Knutson:2013uv}. 

	While most of the short-period gas giants have been detected, the mechanisms under which planets migrate can still be assessed through extended monitoring. Increasing the number of observations for each target star probes for lower mass and longer period planets. The mass detection limit is lowered because increasing the number of observations increases the \snr. Additionally, more widely separated planets can be detected due to a longer observation time baseline. Building up a large population of these long period planets will be useful in determining the occurrence rate of planets at larger separations where they are thought to have formed. These statistically significant occurrence rates can then be used to test and refine population synthesis models that take both disk and gravitational migration mechanisms into account \citep{2013A&A...558A.109A,2013ApJ...775...42I}. Lastly, building up a large sample of long time baseline observations will allow for comparison with other observing methods such as microlensing \citep{2012Natur.481..167C} and direct imaging \citep{Hartung:2013vi,2008SPIE.7014E..41B}. This paper presents the latest 4 planets discovered through the N2K consortium, bringing the total number of exoplanets discovered thus far through the N2K program to 32. 

\section{The N2K Program} \label{sec:n2k}

The N2K target reservoir contains roughly 14,000 stars selected from the \hipp \ Catalog that have $0.4 < B-V < 1.2$, distances closer than 110 parsecs, and $V < 10.5$. Photometric estimates for the temperatures and metallicities of these stars were developed by \citet{2006ApJ...638.1004A}. The reservoir star sample was then ranked according to these metallicity estimates. The N2K program had a very targeted strategy for rapid detection: a set of stars were observed for three or four (nearly) consecutive nights to search for short-period radial velocity variations consistent with orbiting hot Jupiters. Simulations showed that with this observing strategy 90\% of exoplanets with \msini \ $> 0.5$ \mjup\ and orbital periods shorter than 14 days would exhibit 20 \ms\ scatter in the RV measurements \citep{2005ApJ...620..481F}. Stars showing scatter greater than 10 \ms\ were followed up with additional 
observations and stars with low \rms \ scatter were retired to the database. The most obvious short period gas giant planets were detected first; however, monitoring continues on the N2K sample for longer period and multi-planet systems. To date about 560 stars have been observed at Keck as part of the N2K survey, and nearly three dozen planet detections (listed in \autoref{tab:n2k}) have been published from the project, with more emerging planet candidates.  

\section{Data Analysis} 
\label{sec:anal}

High resolution (R$\approx$55,000) spectroscopic observations of the stars discussed in this paper -- HD~5319, HD~10442, HD~75784, and HD~11506 -- were made using Keck HIRES \citep{1994SPIE.2198..362V}. Typical signal-to-noise ratios of our observations were about 150 per pixel. For each star, at least one high resolution observation was taken without the iodine (I$_{2}$) cell in the optical path. From these non-I$_{2}$ spectra, stellar parameters (\teff, [Fe/H], \logg~and \vsini, and elemental abundances for Na, Si, Ti, Fe and Ni) were derived using the LTE spectral synthesis analysis software \textit{Spectroscopy Made Easy} (SME) \citep{1996A&AS..118..595V, 2005ApJS..159..141V}. After generating an initial synthetic model, if parallax measurements were available, we iterated between the Y$^{2}$ isochrones \citep{Demarque:2004p3732} and SME model as described by \citet{2009ApJ...702..989V} until agreement in the surface gravity converged to 0.001 dex.  The stellar mass, luminosity and ages that we present in the following sections were from the Y$^{2}$ isochrones \citep{Demarque:2004p3732}, where bolometric luminosity corrections were adopted from \citet{2003AJ....126..778V}.

The HIRES spectral format includes the \caii\ lines, which are valuable because line core emission in the Ca II lines is a good indicator of chromospheric activity \citep{1984ApJ...279..763N}. This is important for detecting planets via the radial velocity method since chromospheric activity is correlated with increased magnetic fields in stellar photospheres, which drive phenomena like the suppression of convection, stellar spots, and long term activity cycles \citep{1997ApJ...485..319S, 2000ApJ...534L.105S, 2010A&A...511A..54S, 2011arXiv1107.5325L, 2011A&A...527A..82D}. All of these phenomena produce line profile variations that can be misinterpreted as Doppler shifts of the star. These sources of Doppler measurement errors are often combined into one term called stellar ``jitter". \citet{2010ApJ...725..875I} measured emission in the \caii\ line cores to derive \shk\ values and $\log{R'_{HK}}$ (the ratio of emission in the core of the \caii\ lines to the surrounding continuum). They have estimated astrophysical jitter measurements as a function of \bv\ color, luminosity class, and excess \shk\ values, and we have adopted those stellar jitter estimates for the stars in this paper.

Prior to taking Doppler observations, a high resolution (R$\approx$1,000,000), high \snr \ ($\approx$1000) spectrum of an iodine cell was obtained with a Fourier Transform Spectrograph (FTS). This FTS scan was then used in determining Doppler shift measurements with a forward-modeling process \citep{1996PASP..108..500B}. First, the intrinsic stellar spectrum (ISS) was obtained for each star by deconvolving a high resolution, high signal-to-noise non-I$_{2}$ spectrum to remove the spectral line spread function (SLSF), which is sometimes referred to as the point spread function. For all subsequent observations, the iodine cell was placed in the light path to imprint a dense ${\rm I_2}$ absorption spectrum on the stellar spectrum. The iodine lines were used to provide wavelength calibration and to model the SLSF for our observations. Finally, we multiplied the ISS and FTS ${\rm I_2}$ spectra and convolved the product with a SLSF sum of Gaussians model to match each program observation \citep{1995PASP..107..966V}. The modeling process was driven by a Levenberg-Marquardt algorithm, and the free parameters included the Doppler shift, the wavelength solution and the SLSF parameters.  
   
The time series radial velocity data were then analyzed and fit with Keplerian models using \textit{Keplerian Fitting Made Easy\footnotemark} (\kfme) \citep{2012ApJ...744....4G}. This graphical user interface was written in the Interactive Data Language (IDL) as a widget application. Multiple planets in each system can be fit either simultaneously or sequentially. \kfme~includes built in statistical analysis tools, such as periodogram false alarm probability (FAP) and Keplerian FAP tests \citep{2007ApJ...657..533W, 2007ApJ...670..833J, 2009ApJ...696...75H}. Within \kfme, orbital parameter confidence levels can be determined using a bootstrap Monte Carlo method \citep{Press1992, 2005ApJ...619..570M}.

\footnotetext{available at: \url{http:mattgiguere.github.io/KFME}}

A Bayesian approach was also taken to analyze the time series radial velocity measurements for each star. Each set of radial velocity measurements was fit with a Differential Evolution Markov Chain Monte Carlo algorithm. Additionally, for the multi-planet systems dynamical stability was taken into account through n-body integrations, where solutions with close-encounters were rejected \citep{2011AJ....141...16J, Nelson:2013ub}.
 
\capstartfalse
\begin{deluxetable*}{lcccc} 
\tablecaption{Stellar Parameters \label{tab:stellar}} 
\tablewidth{0pt} 
\tablehead{ & \colhead{HD~5319} & \colhead{HD~10442} & \colhead{HD~11506} 
& \colhead{HD~75784}    } 
\startdata 
Spectral type      & K3 IV      & K2 IV        & G0 V       & K3 IV                     \\
$V$        & \vmagfive       & \vmagten       & \vmageleven    & \vmagseven      \\ 
$M_V$    & \absmagfive  & \absmagten   & \absmageleven   & \absmagseven          \\
$B-V$     & \bvfive            & \bvten           & \bveleven            & \bvseven                \\ 
BC          & \bcfive        & \bcten       & \bceleven          & \bcseven           \\
Distance (pc)   & \distfive (\distuncfive)    & \distten  \distuncten    & \disteleven (\distunceleven)   & \distseven
(\distuncseven)  \\ 
$T_{\rm eff}$ (K) & \tefffive (\teffuncfive)    & \tefften (\teffuncten)    & \teffeleven (\teffunceleven)   & \teffseven (\teffuncseven)   \\ 
\logg     & \loggfive  (\logguncfive)    & \loggten (\logguncten)    & \loggeleven (\loggunceleven)   & \loggseven (\logguncseven)   \\ 
${\rm [Fe/H]}$         & \fehfive (\fehuncfive)   & \fehten (\fehuncten)      & \feheleven (\fehunceleven)     &
\fehseven (\fehuncseven)     \\
$M_{\star}$ (\msun)    & \mstfive (\mstuncfive)      & \mstten
(\mstuncten)      & \msteleven (\mstunceleven)     & \mstseven
(\mstuncseven)       \\ 
$R_{\star}$ (\rsun)    & \rstfive (\rstuncfive)      & \rstten \rstuncten     &
\rsteleven (\rstunceleven)     & \rstseven (\rstuncseven)     \\ 
$L_{\star}$ (\lsun)    & \lstfive(\lstuncfive)       & \lstten \lstuncten       &
\lsteleven(\lstunceleven)      &  \lstseven(\lstuncseven)     \\ 
Age (Gyr)              & \agefive 
(\ageuncfive)     & \ageten  \ageuncten     & \ageeleven 
(\ageunceleven)    &  \ageseven  (\ageuncseven)    \\ 
log R'$_{HK}$          & \rhkfive                  
 & \rhkten                   & \rhkeleven                     &
\rhkseven                  \\ 
\enddata 
\end{deluxetable*}
\capstarttrue

\

 \section{HD~5319} 
 \label{sec:hd531} 
 
 \subsection{Stellar Characteristics}
\label{sec:531stel}

Based on measurements from the original \citep{esa97} and the revised \citep{2007A&A...474..653V} \hipp \ Catalog HD~5319 (HIP~4297) is at a distance of \distfive \ $\pm$ \distuncfive \ pc. We adopted the \textit{Hipparcos} V-band magnitude and color of V = \vmagfive \ and \bv\ = \bvfive, applied a bolometric correction of \bcfive\, and calculated the absolute visual magnitude, $M_{V} = $\absmagfive.

An iodine-free ``template" spectrum of HD~5319 was analyzed by iterating SME models with Y$^{2}$ isochrones to derive the following stellar parameters:  \teff~=  \tefffive \ $\pm$ \teffuncfive \ K,  [Fe/H] =
\fehfive \ $\pm$ \fehuncfive \ dex, and \logg~=  \loggfive \ $\pm$ \logguncfive. The isochrone analysis also yields a stellar mass  of \mstfive \ $\pm$ 0.11 \msun, an age of  \agefive \ $\pm$ \ageuncfive \ Gyr, a stellar radius of \rstfive \ $\pm$ \rstuncfive \ R$_{\sun}$, and a luminosity of \lstfive \ $\pm$ \lstuncfive \ L$_{\sun}$. 

SIMBAD and \hipp \ have this star listed as a G5 subgiant; however, based on both our \sme \ results and the \hipp \ B-V measurement, this star most closely resembles a \spectypefive \ \lumclassfive.  HD~5319 has low chromospheric activity with $\log{R'_{HK}} = \ $\rhkfive~and an estimated stellar jitter of \jittervho \ \ms. The stellar properties of HD~5319 are summarized in \autoref{tab:stellar}.

\begin{figure}[h]
\epsfig{file=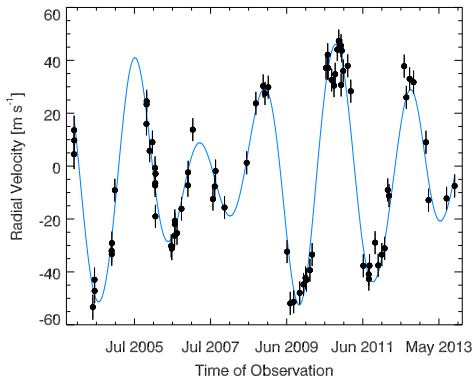,width=1.\linewidth}
\caption{Radial velocity measurements of HD~5319 with associated errors in black.
The best-fit Keplerian model for the two-planet fit is superimposed in blue.}
\label{figHD5319both} \end{figure}

\subsection{Doppler Observations \& Orbital Solution} 
\label{sec:5319orb} 

Based on the first 30 observations of HD~5319 over a time baseline of 3 years, \citet{2007ApJ...670.1391R} announced the discovery of HD~5319~b. Now with a total of \nobsvho \ observations and a time baseline spanning almost 10 years, an additional longer period planetary companion has been confidently detected. The best-fit double Keplerian model yields a slightly revised period for the inner planet of \orbperbvho~$\pm$ \orbperuncbvho\ days with an eccentricity of \orbebvho~$\pm$ \orbeuncbvho. Adopting a stellar mass of \mstfive \ \msun, we derive a planet mass of \orbmbvho~$\pm$ \orbmuncbvho \ \mjup. The newly-discovered outer planet has an orbital period of \orbpercvho~$\pm$ \orbperunccvho\ days, an eccentricity of \orbecvho~$\pm$ \orbeunccvho, and an inferred planet mass of \msini\ = \orbmcvho\ $\pm$ \orbmunccvho\ \mjup. The \rms \ to the two-planet fit is \rmsvho\ \ms. Adding the jitter estimate of \jittervho \ \ms \ from \citet{2010ApJ...725..875I} in quadrature with the formal Doppler errors yields a \chinu \ of \chinuvho. The 81 radial velocities of HD~5319 are shown in black in \autoref{figHD5319both}. 

\begin{figure}[h]
\epsfig{file=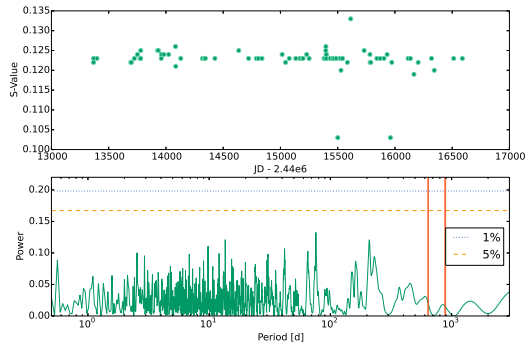,width=1.\linewidth}
\caption{S-Value measurements for each observation of HD~5319 are shown in the top panel and the corresponding GLS periodogram is shown in the bottom panel. Superimposed orange-dashed and blue-dotted horizontal lines indicate the 5\% and 1\% FAP levels, respectively. There is no significant power in the S-values at any period. The orbital periods of the two planets are shown as red vertical bars in the periodogram to highlight that the two planetary signals do not correspond with signals in the S-values.}
\label{figHD5319svals} \end{figure}

One cause for concern is that one (or both) of these signals could be due to the magnetic cycle of the star masquerading as a long-period Keplerian signal. To address these concerns, we performed a generalized Lomb-Scargle periodogram analysis of the S$_{HK}$ values (or simply S-values). We see no signs of magnetic variability in HD~5319. \autoref{figHD5319svals} shows the S-value time series in the top panel and the associated GLS periodogram in the bottom panel. Superimposed in the bottom panel are horizontal lines indicating FAP levels of 5\% (orange-dashed) and 1\% (blue-dotted), which were determined by 1000 bootstrap resamplings of the data \citep{Ivezic:2013ti}, and red vertical lines indicating the orbital periods of the planets. Since there is no peak in the periodogram above the 5\% FAP line, there is no signal in the S-value measurements above 95\% confidence.

\begin{figure}[t]
\epsfig{file=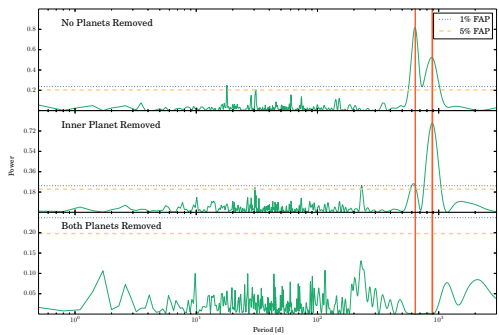,width=1.\linewidth,clip=}
\caption{
(Top) Periodogram of RV measurements of HD~5319 demonstrating the strong periodic signal indicative of a planetary
companion with a $\approx$640 day orbital period. The orange-dashed and blue-dotted horizontal lines show the 5\% and 1\% FAP thresholds, respectively. (Middle) Same as the top with the exception that the Keplerian model for the inner planet has been subtracted. The peak with the highest power corresponds to a period of $\approx$900 days, which motivated fitting for a second planet. (Bottom) Periodogram of the residuals of a two-planet model. The two red vertical lines show the best-fit orbital periods for the two planets. There is no significant power remaining and therefore fitting for additional planets is not warranted.} 
\label{figHD5319periodograms}
\end{figure}

\begin{figure}[hb]
\epsfig{file=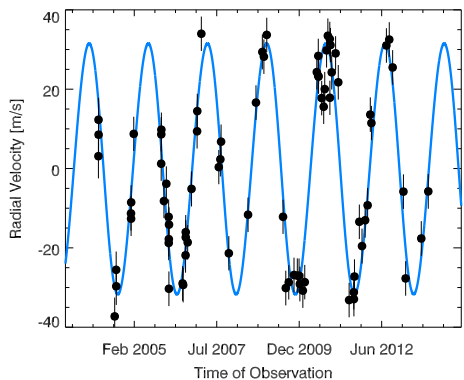,width=1.\linewidth,clip=}
\caption{RV measurements of HD~5319 with the superimposed Keplerian
model for the inner planet, which has an orbital period of \orbperbvho\
days. The Keplerian model for the outer planet has been subtracted.}
\label{figHD5319innerRVS} \end{figure}

Periodogram analysis of the radial velocity measurements shown in the top panel of \autoref{figHD5319periodograms} reveals the dominant signal of the inner planet announced by \citet{2007ApJ...670.1391R} peaking at $\sim$625 days. Also shown in \autoref{figHD5319periodograms} are the 5\% and 1\% FAP levels with orange dashed and blue dotted horizontal lines, respectively. Fitting the 625-day signal using  \kfme \ results in a \chinu \ of 9.3, motivating further inspection. Periodogram analysis of the single-planet fit residuals (middle panel of \autoref{figHD5319periodograms}) reveals an additional signal with a period of 909 days and a FAP  $<$ 0.1\%. Including an additional planet in the Keplerian model results in a significantly lower \chinu \ of \chinuvho. Residual periodogram analysis to the two-planet solution (bottom panel of \autoref{figHD5319periodograms}) reveals no additional signals with significant power indicating any additional planets that may be orbiting HD~5319 are currently below our detection capabilities. 

\begin{figure}[h]
\epsfig{file=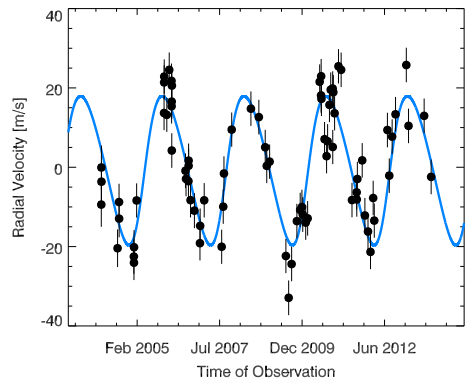,width=1.\linewidth,clip=}
\caption{RV measurements of HD~5319 after subtracting the best-fit
model for the inner planet. The \orbpercvho\ day Keplerian model for the outer
planet is superimposed with a solid blue line. }
\label{figHD5319outer} \end{figure}

The Keplerian models and radial velocity measurements for each of the two planets orbiting HD~5319 have been broken up into two figures to show the contributions and phase coverage of each planet. \autoref{figHD5319innerRVS} shows the residuals after subtracting the Keplerian model for the outer planet from the RV measurements. In other words, it is the radial velocity contribution from just the inner planet. The Keplerian model for just the inner planet is superimposed in blue. This shows both excellent phase coverage and that a Keplerian model accurately describes the data for the inner planet. Similar to \autoref{figHD5319innerRVS}, \autoref{figHD5319outer} shows the contributions and phase coverage of just the outer planet orbiting HD~5319 obtained by subtracting the Keplerian model of the inner planet. Again, there is excellent phase coverage and the Keplerian model for the outer planet accurately describes the data. The orbital parameters for the two planets detected orbiting HD~5319 are summarized in \autoref{tab:orb531}.

\capstartfalse
\begin{deluxetable}{lrr} 
\tablecaption{Orbital Parameters for the HD~5319 System \label{tab:orb531}}
\tablewidth{0pt} 
\tablehead{ \colhead{Parameter}  & \colhead{HD 5319~b} &
\colhead{HD 5319~c}
} 
\startdata 
P(d) & \orbperbvho \ $\pm$ \orbperuncbvho & \orbpercvho \ $\pm$ \orbperunccvho    \\ 
T$_{P}$(JD) & \orbtpbvho \ $\pm$ \orbtpuncbvho& \orbtpcvho \ $\pm$ \orbtpunccvho \\ 
e & \orbebvho \ $\pm$ \orbeuncbvho & \orbecvho \ $\pm$ \orbeunccvho\\ 
$\omega$ & \orbombvho \ $\pm$ \orbomuncbvho & \orbomcvho \ $\pm$ \orbomunccvho \\ 
K  (m s$^{-1}$) & \orbkbvho \ $\pm$ \orbkuncbvho & \orbkcvho \ $\pm$ \orbkunccvho \\ 
a (au) & \orbaplbvho \ $\pm$ \orbapluncbvho & \orbaplcvho \ $\pm$ \orbaplunccvho \\ 
M $\sin{i}$ (\mjup) & \orbmbvho \ $\pm$ \orbmuncbvho & \orbmcvho \ $\pm$ \orbmunccvho \\
N$_{\textrm{obs}}$   &   \nobsvho \\
Jitter (m s$^{-1}$)      & \jittervho \\ 
\rms~(m s$^{-1}$) & \rmsvho \\
$\chi_{\nu}^{2}$ & \chinuvho \\ 
\enddata 
\end{deluxetable}
\capstarttrue

\begin{figure}[hb]
\epsfig{file=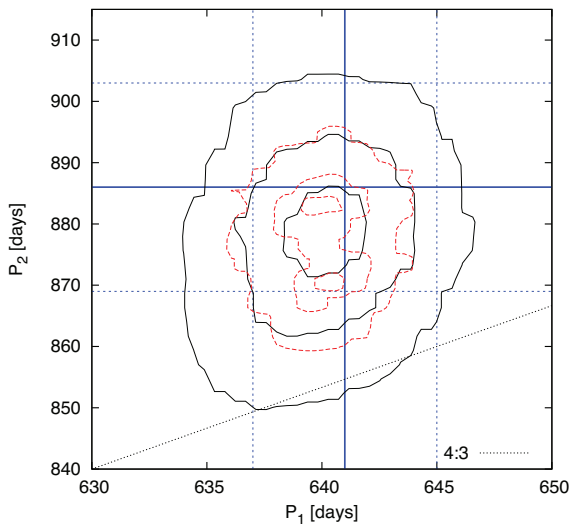,width=1.\linewidth,angle=0,clip=}
\caption{Levenberg-Marquardt, Keplerian MCMC and N-body MCMC solutions for the orbital periods for the 2
planets orbiting HD~5319. The black lines are the 25\%, 1-$\sigma$ and 2-$\sigma$ contours from a 
Keplerian MCMC analysis. The red contours are the same as the black, except dynamical
stability has been taken into account through N-body simulations. The blue lines show the best-fit solution (solid)
and 2-$\sigma$ uncertainties (dashed) using a Levenberg-Marquardt approach. To guide the eye, we have plotted the 4:3 period ratio as a diagonal dotted line.}
\label{figHD5319_demcmc} 
\end{figure}

To further increase confidence in our two-planet interpretation, we searched for a linear correlation between the S-values and RV measurements. The Pearson correlation coefficient ($\rho$) between the raw RV measurements and the S-values was -0.11. To quantify the lack of significance of this anti-correlation we created a distribution of $\rho$ by randomly sampling from the RVs with replacement (i.e., bootstrapping) 10,000 times. A p-value was then determined by counting the fraction of randomly sampled data sets that had a $|\rho|$ greater than our initial $\rho$ of the unscrambled data set. In this case our p-value was 0.33, which implies our measured $\rho$ does not differ significantly from the null hypothesis of $\rho$=0. We performed this same test with the residual velocity measurements: after subtracting the dominant (previously-published inner planet) 641-day signal we calculated a $\rho$ of -0.17 with a corresponding p-value of 0.16, again showing no significant linear correlation between the two parameters. Subtracting the Keplerian model of the outer planet and repeating this analysis resulted in a $\rho$ of 0 with a p-value of 1. Similarly, the result for the residuals to the fit for both planets was $\rho$=-0.10 with a p-value of 0.41. All of these tests were consistent with the null hypothesis, meaning there's no correlation between the two parameters and reaffirming our two-planet interpretation. 

As described at the end of \S\ref{sec:anal}, in addition to fitting the RV measurements with a Levenberg-Marquardt Least Squares Minimization scheme with \kfme, a Bayesian approach was taken to analyze the data using the RUN DMC algorithm \citep{Nelson:2013ub}. First, the radial velocity measurements were fitted with a double-Keplerian model using DEMCMC without taking dynamical stability into account. The resulting distribution of periods for the inner and outer planets are shown in black in \autoref{figHD5319_demcmc} with 25\%, 1-$\sigma$ and 2-$\sigma$ confidence level contours. The blue solid lines are the best-fit solutions from \kfme \ discussed earlier with the 2-$\sigma$~confidence levels shown as blue dashed lines. DEMCMC analysis produces a median solution that is consistent with the \kfme \ result. The median periods from Keplerian MCMC analysis for the inner and outer planets with 1-$\sigma$ confidence levels are $ 640.1^{+2.4}_{-2.5}$ and $ 878^{+10}_{-11}$ days, respectively. The inferred minimum masses for the inner and outer planets are $1.68 \pm 0.07$ and $1.03 \pm 0.09$ \mjup, respectively. 

\begin{figure}[h]
\epsfig{file=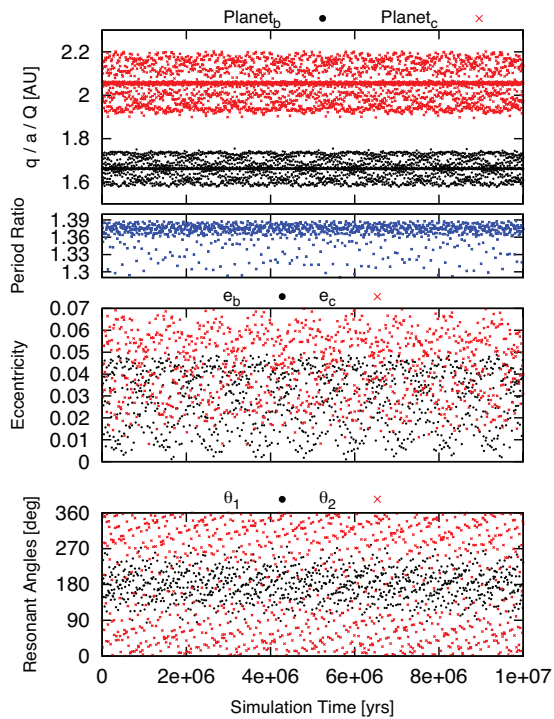,width=1.\linewidth,angle=0,clip=}
\caption{One of the long-term dynamical evolution simulations where the HD~5319 system was stable over the entire $10^{7}$ yr integration period. Shown are the orbital semi-major axes ($a_b\,\&\,a_c$), period ratio, eccentricities ($e_b\,\&\,e_c$), and resonant angles ($\theta_{1}$ and $\theta_2$) as functions of time for the two planets orbiting HD~5319. We see that the system stably librates within the 4:3 resonance.}
\label{figHD5319_stable} 
\end{figure}

When a 100-year dynamical stability constraint is included with the RUN DMC algorithm, the majority of solutions are concentrated into the same region of parameter space as the Keplerian MCMC and \kfme \ results, with median values for the orbital periods of the inner and outer planets of $640.1 \pm 1.2$ and $878^{+6}_{-9}$ days, respectively. The only significant difference between the Keplerian MCMC and RUN DMC solutions is that when the 100-year dynamical stability is enforced the orbital period uncertainty decreases. These RUN DMC results are superimposed in red long-dashed contours showing the 25\%, 1-$\sigma$ and 2-$\sigma$ levels in \autoref{figHD5319_demcmc}. 

	Simulations testing the dynamical stability over longer periods (10$^{7}$ years) were carried out using \texttt{MERCURY} \citep{Chambers:1999p3144}. Most of the realizations were unstable; however, several remained stable over the duration of the simulations. Of the realizations that were stable, all of them exhibited libration. While all three fitting methods resulted in best-fit period ratios that were slightly higher than 4:3, the best-fit solution only reflects our instantaneous ``snapshot" of the system. Since all the long-term simulations that were stable exhibited libration, the long-term averaged orbital period ratio may be 4:3, which would put this system in the 4:3 mean motion resonance and not just close to it. In \autoref{figHD5319_stable} we provide an example of one of the stable solutions which occupy the 4:3 resonance, illustrating the oscillations in orbital elements as a function of time, as well as the evidence for libration in the resonant angles, $\theta_1$ and $\theta_2$. The resonant angles are defined as 
	
\begin{eqnarray}
\theta_1 = 4\left(\lambda_2 - \varpi_2 \right) - 3\left(\lambda_1 - \varpi_1 \right) + 3\left(\varpi_2 - \varpi_1 \right) \\
\theta_2 = 4\left(\lambda_2 - \varpi_2 \right) - 3\left(\lambda_1 - \varpi_1 \right) + 4\left(\varpi_2 - \varpi_1 \right)
\end{eqnarray}

where $\lambda_i$ and $\varpi_i$ are the mean longitude and longitude of periapse of the $i^{th}$ planet, respectively. This demonstrates that this system is stably librating within the 4:3 resonance. We note that the planetary eccentricities are highly oscillatory, with $e_b$ in this example frequently returning to an approximately circular ($e_b=0$) state.

\LongTables
\capstartfalse
\begin{deluxetable}{rrrr}
\tablecaption{RV Measurements of HD~5319 \label{tab:rvs5319}}
\tablewidth{0pt}
\tablehead{\colhead{JD}  & \colhead{RV}  
& \colhead{$\sigma_{RV}$} & \colhead{$S_{HK}$}   \\
\colhead{-2440000}  & \colhead{(\ms)}   & \colhead{(\ms)}  } 
\startdata
13014.7556  &  4.50  &  3.70 &   \\  
13015.7606  &  13.58  &  3.61 &   \\  
13016.7651  &  9.72  &  2.99 &   \\  
13191.1101  &  -53.30  &  2.22 &   \\  
13207.0754  &  -42.94  &  1.90 &   \\  
13208.0656  &  -47.18  &  2.12 &   \\  
13367.7078  &  -32.05  &  1.13 & 0.122  \\  
13368.7160  &  -33.30  &  1.22 & 0.123  \\  
13369.7253  &  -29.11  &  1.03 & 0.123  \\  
13397.7194  &  -9.11  &  1.00 & 0.123  \\  
13694.7659  &  15.97  &  1.02 & 0.122  \\  
13695.7711  &  23.36  &  0.98 & 0.122  \\  
13696.7462  &  24.52  &  1.00 & 0.122  \\  
13724.7791  &  5.79  &  0.91 & 0.123  \\  
13750.7363  &  9.04  &  1.28 & 0.124  \\  
13775.7200  &  -0.69  &  1.24 & 0.123  \\  
13776.7061  &  -7.33  &  1.19 & 0.123  \\  
13777.7208  &  -6.42  &  1.26 & 0.123  \\  
13778.7168  &  -19.00  &  1.20 & 0.123  \\  
13779.7412  &  -2.90  &  1.17 & 0.125  \\  
13927.0485  &  -30.18  &  1.22 & 0.125  \\  
13933.0449  &  -31.17  &  1.24 & 0.125  \\  
13959.0917  &  -26.29  &  1.27 & 0.123  \\  
13961.0367  &  -21.96  &  1.01 & 0.124  \\  
13961.0402  &  -20.66  &  1.00 & 0.124  \\  
13981.9060  &  -25.36  &  1.18 & 0.124  \\  
14023.7760  &  -16.13  &  1.36 & 0.124  \\  
14083.8337  &  -7.27  &  1.13 & 0.126  \\  
14085.9027  &  -2.36  &  1.22 & 0.121  \\  
14129.7746  &  13.80  &  1.08 & 0.123  \\  
14319.0740  &  -12.47  &  0.98 & 0.123  \\  
14336.0524  &  -7.68  &  1.01 & 0.123  \\  
14343.9381  &  -1.85  &  1.06 & 0.123  \\  
14427.9088  &  -15.65  &  0.98 & 0.123  \\  
14636.0956  &  1.25  &  1.12 & 0.125  \\  
14721.9810  &  23.72  &  1.20 & 0.123  \\  
14790.8874  &  30.29  &  1.28 & 0.123  \\  
14807.8193  &  27.46  &  1.13 & 0.123  \\  
14838.7959  &  29.85  &  1.08 & 0.123  \\  
15015.1236  &  -32.33  &  1.01 & 0.124  \\  
15045.0752  &  -51.89  &  1.23 & 0.122  \\  
15077.0750  &  -51.30  &  1.07 & 0.123  \\  
15133.9738  &  -47.91  &  1.15 & 0.123  \\  
15169.8628  &  -44.74  &  1.14 & 0.123  \\  
15188.7784  &  -42.27  &  1.17 & 0.123  \\  
15197.7498  &  -43.02  &  1.22 & 0.123  \\  
15229.7113  &  -39.36  &  1.14 & 0.124  \\  
15250.7109  &  -33.44  &  1.11 & 0.123  \\  
15381.1261  &  37.15  &  1.24 & 0.123  \\  
15396.1025  &  37.10  &  1.21 & 0.125  \\  
15397.0565  &  42.13  &  1.09 & 0.126  \\  
15400.0755  &  37.05  &  1.09 & 0.124  \\  
15434.0867  &  32.58  &  1.10 & 0.123  \\  
15455.9742  &  30.44  &  1.10 & 0.123  \\  
15467.0374  &  34.77  &  1.05 & 0.123  \\  
15487.0331  &  44.07  &  1.08 & 0.123  \\  
15500.8621  &  47.28  &  1.19 & 0.103  \\  
15521.8665  &  45.55  &  1.10 & 0.123  \\  
15522.8818  &  30.67  &  1.00 & 0.123  \\  
15528.8672  &  43.64  &  1.11 & 0.120  \\  
15542.8488  &  35.99  &  1.04 & 0.123  \\  
15584.7044  &  37.93  &  1.04 & 0.122  \\  
15613.7048  &  28.33  &  1.15 & 0.133  \\  
15731.1069  &  -37.67  &  1.12 & 0.125  \\  
15782.1153  &  -40.90  &  1.04 & 0.124  \\  
15783.1368  &  -42.71  &  1.14 & 0.122  \\  
15789.1394  &  -37.63  &  1.10 & 0.122  \\  
15841.9586  &  -28.92  &  1.13 & 0.123  \\  
15871.0018  &  -37.54  &  1.35 & 0.123  \\  
15904.7652  &  -33.46  &  1.28 & 0.123  \\  
15931.7529  &  -31.07  &  1.15 & 0.124  \\  
15960.7076  &  -9.01  &  1.14 & 0.103  \\  
15972.7134  &  -11.19  &  1.03 & 0.122  \\  
16115.1335  &  37.79  &  1.17 & 0.123  \\  
16135.1396  &  26.00  &  1.13 & 0.123  \\  
16166.1509  &  33.00  &  1.22 & 0.119  \\  
16202.9785  &  31.74  &  1.19 & 0.122  \\  
16319.7103  &  9.01  &  1.17 & 0.123  \\  
16343.7140  &  -12.85  &  1.16 & 0.120  \\  
16513.1415  &  -12.23  &  1.23 & 0.123  \\  
16588.9789  &  -7.49  &  1.22 & 0.123  \\  
\enddata
\end{deluxetable}
\capstarttrue

	There have been several systems near or in 4:3 mean motion resonances discovered by the \textit{Kepler} Mission \citep{2011ApJ...728..117B, 2011ApJS..197....8L, 2013ApJS..204...24B} and one other system (HD~200964) discovered by the radial velocity method \citep{2011AJ....141...16J}. The HD~200694 and HD~5319 systems are qualitatively different than the \textit{Kepler} systems, in that the orbiting planets in these RV-detected systems have much longer orbital periods and are much more massive. A recent study by \citet{2012MNRAS.426..187R} has tested a variety of formation and migration mechanisms attempting to recreate the observed distribution of planets in or near the 4:3 resonance. They found that while they could recreate the low mass \textit{Kepler} systems, they could not reproduce gas giants in a 4:3 resonance. They tried several mechanisms for forming such a system: convergent migration, scattering and simultaneous damping, and in situ formation. All simulations either failed to create a system near a 4:3 resonance or required highly tuned initial conditions that produced 1:1 resonances three times more frequently, none of which have been observed. Although only two massive systems are known to be near the 4:3 resonance, \citet{2012MNRAS.426..187R} conclude that the observed fraction of such systems is too high to explain with traditional formation mechanisms. They suggest two additional mechanisms that have not yet been investigated: resonant chain breaking and chaotic migration. However, the exact mechanism behind the formation of the HD~200964 and HD~5319 planetary systems is still a puzzle to be solved.

\begin{figure}[h]
\epsfig{file=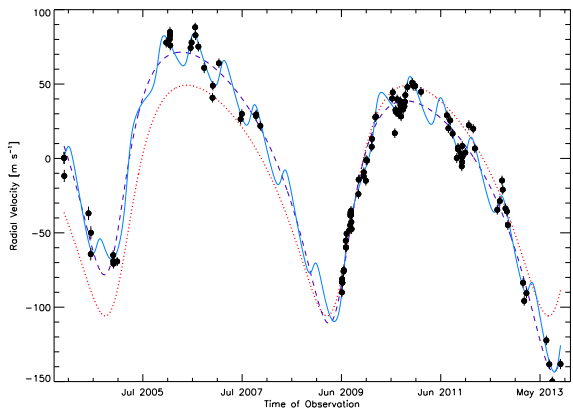,width=1.\linewidth}
\caption{Radial velocity measurements for HD~11506 (black). Superimposed are the
theoretical models for one planet with no linear trend (red dotted), one planet
with a linear trend (purple dashed), and two planets with a linear trend (blue solid).
} \label{figHD11506rvs} 
\end{figure}

\section{HD~11506} \label{sec:hd115}

\subsection{Stellar Characteristics}
\label{sec:115stel}

HD~11506 (HIP~8770) is an early G dwarf star observed as part of the original N2K survey. The trigonometric parallax listed in the \hipp \ catalog is 19.34 \ $\pm$ 0.58 mas, which corresponds to a distance 51.7 \ $\pm$ 0.6 pc. Combined with the Johnson V magnitude of \vmageleven \ also listed in the \hipp \ catalog, we calculate an absolute visual magnitude of \absmageleven. Iteration between \sme \ and the \yy \ isochrones as described in \S \ref{sec:anal} results in a best-fit metallicity of \feheleven \ $\pm$ \fehunceleven; a surface gravity of \loggeleven \ $\pm$  \loggunceleven; and an effective temperature of \teffeleven \ $\pm$ \teffunceleven \ K. From iteration with the isochrones the stellar mass converges to \msteleven \ $\pm$ \mstunceleven \ \msun \ with a stellar radius of \rsteleven \ $\pm$ \rstunceleven \ \rsun, a stellar luminosity of \lsteleven \ $\pm$ \lstunceleven \ \lsun, and an age of \ageeleven \ $\pm$ \ageunceleven \ Gyr. The stellar parameters for HD~11506 are summarized in \autoref{tab:stellar}. 

\subsection{Doppler Observations \& Orbital Solution} 
\label{sec:115orb}

With 3.5 years of data accumulated, \citet{2007ApJ...669.1336F} announced the discovery of a planet orbiting HD~11506 with a period of 1405 days. They noted several remaining peaks in a periodogram of the residuals, including a peak at 170 days; however, they cautioned that more data were required to evaluate the second signal. \citet{2009A&A...496L..13T} carried out an extensive Bayesian analysis claiming that the second planet did indeed exist with an orbital period of 170.5$^{+3.3}_{-6.2}$ days. However, their Bayesian analysis is roughly 48 $\sigma$ from the period we obtain with our extended data set.

\begin{figure} 
\epsfig{file=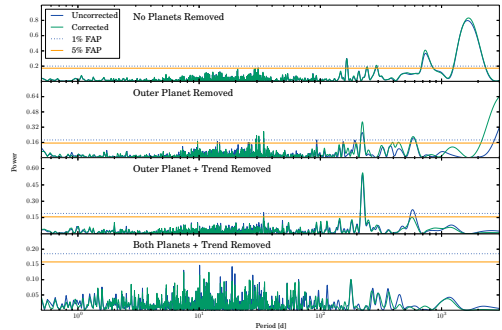,width=1.\linewidth}
\caption{Periodograms of the RV measurements of HD~11506 and the residuals after subtracting Keplerian and linear models. The green curve shows the power in the magnetic-signal-corrected time series and the blue curve shows the power in the uncorrected time series.} 
\label{figHD11506pergs} 
\end{figure}

Since the initial discovery paper by \citet{2007ApJ...669.1336F}, HD~11506 has been observed an additional 87 times over the past 6 years. After subtracting the best-fit $\sim$1600-day Keplerian model (represented by a red-dotted line in \autoref{figHD11506rvs}) from the RV measurements, the extended data set shows a clear long-period signal that is the dominant power in the periodogram of the residuals. This is shown in the second panel from the top in \autoref{figHD11506pergs}. The period of this long-period signal is much longer than the time baseline of our observations, and it can be well-approximated as a linear trend. Incorporating this linear term (the purple-dashed line in \autoref{figHD11506rvs}) reduced the \chinu \ from 36 for the single-planet fit to 9 for the single-planet + linear trend fit. Periodogram analysis of the residuals of the single-planet + linear trend model reveals the presence of an additional companion with a 223 day period, which is shown in the second panel from the bottom in \autoref{figHD11506pergs}. The best-fit two-planet + linear trend solution, which has a  \chinu \ of 2.94 and an rms of 5.8 is superimposed in solid blue in \autoref{figHD11506rvs}. Since there is no significant power remaining in the residuals, fitting for additional planets is not warranted.

\begin{figure} [t]
\epsfig{file=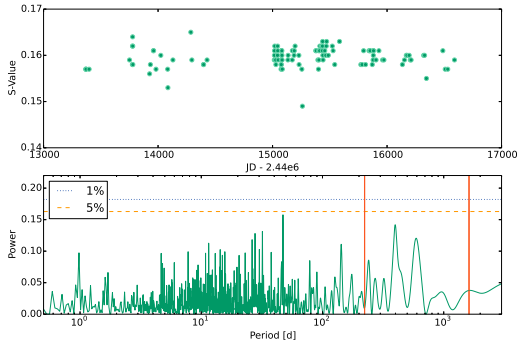,width=1.\linewidth}
\caption{Same as \autoref{figHD5319svals}, but for S-values of HD~11506 observations. Here it can be seen that there is no significant power due to magnetic activity at either of the planetary signals.
} \label{figHD11506svals} 
\end{figure}

Similar to tests performed on the HD~5319 observations, we searched for signs of magnetic activity contributing to the RV measurements. \autoref{figHD11506svals} shows the S-value time series (top panel) and the GLS periodogram of the S-values (bottom panel); the orbital periods of the two planets are superimposed in the bottom panel as red vertical lines. This shows that there is no period with $>$ 95\% confidence at any power, and there are no peaks corresponding to the periods of the planets. Furthermore, periodogram power does not significantly increase towards longer periods, indicating the linear trend is not due to magnetic activity either.

\begin{figure}[h]
\epsfig{file=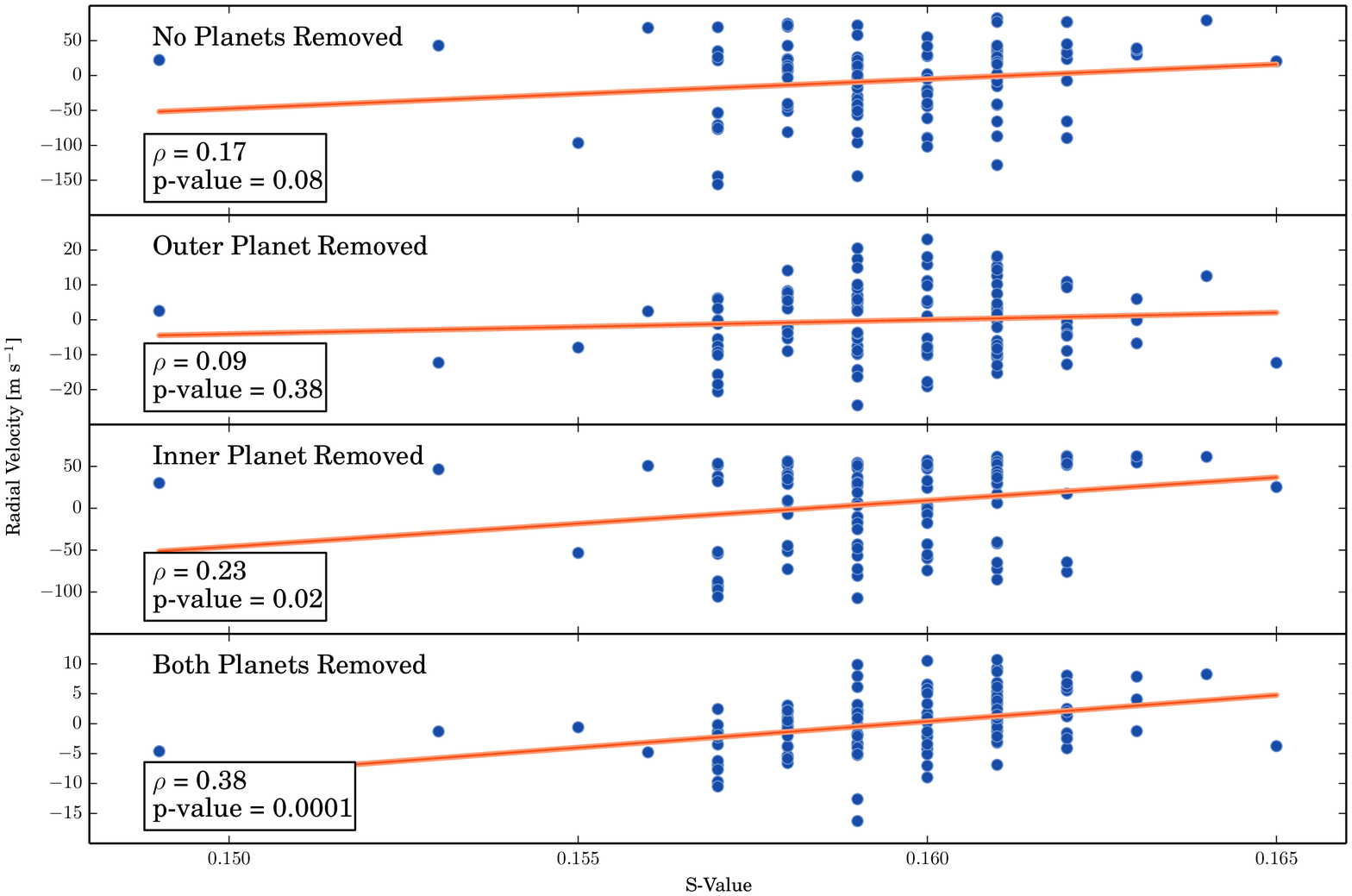,width=1.\linewidth}
\caption{RV Measurements and residuals for HD~11506 as a function of S-value. The top panel shows the raw RV measurements with no planetary Keplerian models subtracted, the middle two panels show the residuals after subtracting each best-fit keplerian model, and the bottom panel shows the residuals after subtracting both Keplerian models and a linear term from the RV data. There is no statistically significant correlation between the velocities and S-values when both planetary signals are present, but subtracting the planetary signals reveals a linear correlation between the residuals and the S-value measurements.} 
\label{figHD11506svalcor} 
\end{figure}

We also looked for correlations between the residual velocity measurements and the S-values as we did with HD~5319. The top panel of \autoref{figHD11506svalcor} shows the RV measurements as a function of S-value for the raw velocity set. As described in \S \ref{sec:5319orb}, we calculated the Pearson correlation coefficient, $\rho$, and its associated p-value for these two parameters, which were 0.17 and 0.08, respectively. This indicates that there is no significant correlation between the ``raw" velocities and the S-values. We then repeated the analysis after subtracting the Keplerian model of the outer planet from the velocity measurements; the results are shown in the second panel from the top in \autoref{figHD11506svalcor}. Again, there is no statistically significant linear correlation between the two parameters. Removing the model for the inner planet without removing the linear trend resulted in $\rho$ = 0.19 with a p-value of 0.05. To check the impact of this marginally significant linear correlation on our two-planet + linear trend interpretation we subtracted the linear model that best-fit the RV-S-value data from the velocity measurements and repeated the Keplerian modeling. This resulted in similar orbital parameters between the magnetic-signal-corrected and non-magnetic-signal-corrected RV measurements, but with an increase in rms. We then subtracted the Keplerian model of the inner planet and linear trend from the velocities and performed the same test on the residuals. Interestingly, a significant correlation between the two parameters emerged (second panel from the bottom). 

\begin{deluxetable}{lrr} 
\tablecaption{Orbital Parameters for the HD~11506 System \label{tab:orb115}}
\tablewidth{0pt} 
\tablehead{ \colhead{Parameter}  & \colhead{HD 11506~b}
& \colhead{HD 11506~c}
} 
\startdata 
P(d) & \orbperboov \ $\pm$ \orbperuncboov & \orbpercoov \ $\pm$ \orbperunccoov    \\ 
T$_{P}$(JD) & \orbtpboov \ $\pm$ \orbtpuncboov& \orbtpcoov \ $\pm$\orbtpunccoov \\ 
e & \orbeboov \ $\pm$ \orbeuncboov & \orbecoov \ $\pm$ \orbeunccoov\\ 
$\omega$ & \orbomboov \ $\pm$ \orbomuncboov & \orbomcoov \ $\pm$ \orbomunccoov \\ 
K  (m s$^{-1}$) & \orbkboov \ $\pm$ \orbkuncboov & \orbkcoov \ $\pm$ \orbkunccoov \\ 
a (au) & \orbaplboov \ $\pm$ \orbapluncboov & \orbaplcoov \ $\pm$ \orbaplunccoov \\ 
M $\sin{i}$  (\mjup) & \orbmboov \ $\pm$ \orbmuncboov & \orbmcoov \ $\pm$ \orbmunccoov \\
\textit{dv/dt} (m s$^{-1}$ yr$^{-1}$) & \dvdtoov \ $\pm$ \dvdtuncoov \\ 
N$_{\textrm{obs}}$   &   \nobsoov \\
Jitter (m s$^{-1}$)      & \jitteroov \\ 
\rms~(m s$^{-1}$) & \rmsoov \\
$\chi_{\nu}^{2}$ & \chinuoov \\ \enddata \end{deluxetable}

Lastly, we subtracted the full two-planet + linear trend model from the velocities and saw a very significant linear correlation between the two parameters ($\rho$ = 0.38, p-value = 0.0001), which can be seen in the bottom panel of \autoref{figHD11506svalcor}. Subtracting the best-fit linear model between these residual RV measurements and S-values from the original RV measurements and repeating the Keplerian analysis resulted in similar orbital parameters; however, after correcting for the magnetic signal the peaks in the periodograms were slightly higher at the periods corresponding to the planetary signals. We were interested to see if subtracting this magnetic signal and refitting would reveal additional planets in the system that were previously below the stellar noise level; however, power in the highest peaks in the residual periodogram decreased after correcting for the magnetic signal and no new signals emerged. 

This Keplerian signal enhancement and noise reduction can be seen in \autoref{figHD11506pergs}, where the uncorrected periodogram power is shown in blue and the magnetic-signal-corrected power is shown in green. Overall, subtracting the magnetic signal reduced the rms by 1.0 \ms (21\%) relative to the uncorrected result. The final refined best-fit solution for the outer planet has an orbital period of \orbperboov \ $\pm$ \orbperuncboov \ days, an eccentricity of \orbeboov \ $\pm$ \orbeuncboov, and a radial velocity semi-amplitude of \orbkboov \ $\pm$ \orbkuncboov \ \ms. Based on these parameters the calculated semi-major axis is \orbaplboov \ $\pm$ \orbapluncboov \ AU and the planet has a minimum mass of \orbmboov \ $\pm$ \orbmuncboov \ \mjup. The inner planet has a best-fit solution with an orbital period of \orbpercoov \ $\pm$ \orbperunccoov \ days, an eccentricity of \orbecoov \ $\pm$ \orbeunccoov, and a semi-amplitude of \orbkcoov \ $\pm$ \orbkunccoov \ \ms. This corresponds to a semi-major axis of \orbaplcoov \ $\pm$ \orbaplunccoov \ AU and a minimum mass of \orbmcoov \ $\pm$ \orbmunccoov \ \mjup. These orbital parameters are summarized in \autoref{tab:orb115} .

\begin{figure}
\epsfig{file=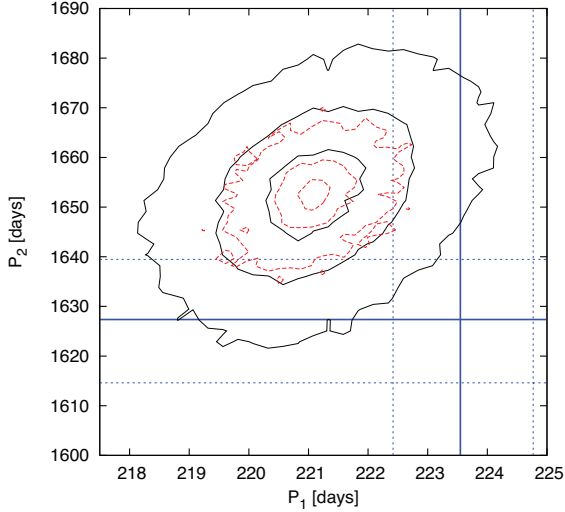,width=1.\linewidth,angle=0,clip=}
\caption{Levenberg-Marquardt, Keplerian MCMC and N-body MCMC solutions for the orbital periods for the 2 planets orbiting HD~11506. The black lines are the 25\%, 1-$\sigma$ and 2-$\sigma$ contours from a Keplerian MCMC analysis. The red contours are the same as the black with the exception that dynamical stability has been taken into account through N-body simulations. The blue lines show the best-fit solution (solid) and 2-$\sigma$ uncertainties (dashed) using the \textit{KFME} Levenberg-Marquardt approach. }
\label{figHD11506_DEMCMC2} 
\end{figure}

The Keplerian DEMCMC analysis shown in black in \autoref{figHD11506_DEMCMC2} resulted in orbital periods for the inner and outer planets of $221.1 \pm 1.1$ and $1653^{+11}_{-12}$ days, and minimum masses of $ 0.50 \pm 0.06$ and $ 5.5 \pm 0.2$ \mjup, respectively. Taking into account n-body interactions over short timescales, the n-body RUN DMC analysis resulted in consistent results with orbital periods of $221.1 \pm 0.4$ and $1653 \pm 4$ days, and minimum masses of $0.48^{0.03}_{-0.04}$ and $ 5.5 \pm 0.1$ for the inner and outer planets, respectively. These are superimposed in \autoref{figHD11506_DEMCMC2} in red along with the \kfme \ results, which are in blue. The resulting period distributions from the DEMCMC and \kfme \ analysis do not lie on top of each other, but there is considerable overlap in the 2-$\sigma$ wings of the distributions. There are a few possible explanations for the separation in the orbital solutions for the two models: the DEMCMC result is the median of the period distributions whereas the \kfme \ solution is the best-fit, the difference in priors in each model plays a role, and the noise is assumed to be normally-distributed in the \kfme \ analysis when it is most likely not exactly Gaussian. 

\capstartfalse
\begin{deluxetable}{rrrr}
\tablecaption{RV Measurements of HD~11506 \label{tab:rvs11506}}
\tablewidth{0pt}
\tablehead{\colhead{JD}  & \colhead{RV}  
& \colhead{$\sigma_{RV}$} & \colhead{$S_{HK}$}   \\
\colhead{-2440000}  & \colhead{(\ms)}   & \colhead{(\ms)}  } 
\startdata
13014.7350  &  0.29  &  2.99 &   \\  
13015.7389  &  0.39  &  2.99 &   \\  
13016.7409  &  -11.77  &  2.99 &   \\  
13191.1220  &  -36.93  &  3.59 &   \\  
13207.1012  &  -64.34  &  3.14 &   \\  
13208.0840  &  -49.96  &  3.32 &   \\  
13368.8378  &  -65.01  &  1.70 & 0.157  \\  
13369.7590  &  -69.06  &  1.48 & 0.157  \\  
13370.7324  &  -70.80  &  1.65 & 0.157  \\  
13397.7301  &  -69.26  &  1.24 & 0.157  \\  
13750.7381  &  77.84  &  1.76 & 0.159  \\  
13775.7285  &  80.20  &  1.83 & 0.158  \\  
13776.7044  &  82.52  &  1.67 & 0.162  \\  
13777.7253  &  85.05  &  1.88 & 0.164  \\  
13778.7186  &  82.70  &  2.03 & 0.162  \\  
13779.7474  &  75.99  &  1.82 & 0.158  \\  
13926.1274  &  74.32  &  1.70 & 0.156  \\  
13933.0906  &  77.90  &  1.76 & 0.158  \\  
13959.1394  &  88.17  &  1.26 & 0.161  \\  
13961.1242  &  82.82  &  1.55 & 0.161  \\  
13981.9826  &  75.16  &  1.67 & 0.157  \\  
14023.9744  &  60.88  &  2.10 & 0.160  \\  
14083.8433  &  40.80  &  1.46 & 0.157  \\  
14085.9211  &  48.89  &  1.47 & 0.153  \\  
14129.7433  &  64.00  &  1.48 & 0.159  \\  
14286.1184  &  26.39  &  2.00 & 0.165  \\  
14295.0948  &  29.94  &  1.47 & 0.159  \\  
14396.8462  &  28.57  &  1.41 & 0.158  \\  
14397.9753  &  29.55  &  1.50 & 0.158  \\  
14427.9127  &  21.88  &  1.50 & 0.159  \\  
15015.1201  &  -90.04  &  1.62 & 0.159  \\  
15016.1145  &  -83.23  &  1.50 & 0.160  \\  
15017.1212  &  -81.06  &  1.53 & 0.161  \\  
15019.1232  &  -83.56  &  1.58 & 0.162  \\  
15027.1128  &  -75.93  &  1.46 & 0.159  \\  
15029.1124  &  -75.08  &  1.54 & 0.158  \\  
15043.1284  &  -55.30  &  1.55 & 0.160  \\  
15044.1334  &  -59.71  &  1.69 & 0.162  \\  
15045.1092  &  -59.90  &  1.58 & 0.161  \\  
15049.1053  &  -50.52  &  1.56 & 0.159  \\  
15074.1001  &  -47.75  &  1.55 & 0.157  \\  
15075.1049  &  -44.87  &  1.58 & 0.158  \\  
15076.0998  &  -38.41  &  1.67 & 0.158  \\  
15077.0913  &  -37.04  &  1.50 & 0.159  \\  
15078.0946  &  -36.01  &  1.50 & 0.161  \\  
15081.1111  &  -35.22  &  1.59 & 0.161  \\  
15082.0962  &  -37.89  &  1.57 & 0.160  \\  
15083.1074  &  -42.78  &  1.48 & 0.159  \\  
15085.0652  &  -47.45  &  1.75 & 0.157  \\  
15133.9969  &  -23.94  &  1.58 & 0.159  \\  
15135.9215  &  -14.21  &  1.59 & 0.160  \\  
15171.8999  &  -12.04  &  1.72 & 0.159  \\  
15172.8826  &  -9.32  &  1.65 & 0.161  \\  
15187.7268  &  -14.97  &  1.64 & 0.160  \\  
15189.8075  &  -1.12  &  1.56 & 0.161  \\  
15196.7897  &  -1.44  &  1.49 & 0.162  \\  
15229.7192  &  7.82  &  1.58 & 0.160  \\  
15231.7230  &  13.15  &  1.63 & 0.158  \\  
15255.7125  &  27.73  &  1.56 & 0.157  \\  
15260.7136  &  28.13  &  1.76 & 0.149  \\  
15377.1272  &  40.29  &  1.53 & 0.161  \\  
15381.1188  &  44.36  &  1.52 & 0.161  \\  
15396.1222  &  16.99  &  1.61 & 0.159  \\  
15401.0718  &  32.53  &  1.53 & 0.157  \\  
15403.1233  &  31.33  &  1.38 & 0.161  \\  
15405.0935  &  32.10  &  1.54 & 0.159  \\  
15411.1124  &  31.23  &  1.57 & 0.161  \\  
15413.0789  &  39.74  &  1.57 & 0.162  \\  
15426.0844  &  29.64  &  1.60 & 0.162  \\  
15435.0957  &  32.51  &  1.57 & 0.161  \\  
15436.0942  &  38.45  &  1.55 & 0.162  \\  
15437.1286  &  35.85  &  1.58 & 0.163  \\  
15439.1314  &  33.75  &  1.40 & 0.160  \\  
15441.1157  &  28.28  &  1.56 & 0.159  \\  
15455.9780  &  32.93  &  1.56 & 0.161  \\  
15465.0604  &  35.01  &  1.61 & 0.160  \\  
15469.0649  &  38.09  &  1.51 & 0.162  \\  
15471.9672  &  42.54  &  1.64 & 0.163  \\  
15487.0061  &  47.90  &  1.56 & 0.160  \\  
15521.8824  &  50.90  &  1.65 & 0.162  \\  
15528.8094  &  48.84  &  1.80 & 0.161  \\  
15542.9508  &  48.69  &  1.65 & 0.158  \\  
15584.7831  &  44.77  &  1.56 & 0.163  \\  
15771.0854  &  29.00  &  1.54 & 0.158  \\  
15782.1102  &  20.24  &  1.50 & 0.158  \\  
15795.1323  &  25.54  &  1.63 & 0.161  \\  
15812.1130  &  16.63  &  1.66 & 0.158  \\  
15842.0209  &  0.25  &  1.81 & 0.161  \\  
15850.9531  &  7.24  &  1.54 & 0.161  \\  
15852.0283  &  6.02  &  1.60 & 0.161  \\  
15870.9983  &  2.92  &  1.78 & 0.158  \\  
15877.9871  &  -5.31  &  1.64 & 0.159  \\  
15879.9690  &  0.77  &  1.56 & 0.160  \\  
15880.8704  &  -2.41  &  1.60 & 0.161  \\  
15881.8258  &  8.39  &  1.55 & 0.161  \\  
15903.7783  &  3.90  &  1.57 & 0.159  \\  
15928.8013  &  22.31  &  1.96 & 0.161  \\  
15960.7459  &  19.97  &  1.46 & 0.159  \\  
15972.7083  &  6.70  &  1.52 & 0.159  \\  
16134.1399  &  -34.52  &  1.60 & 0.158  \\  
16152.1129  &  -28.66  &  1.55 & 0.159  \\  
16168.0626  &  -14.84  &  1.72 & 0.160  \\  
16173.1109  &  -20.99  &  1.58 & 0.160  \\  
16193.0920  &  -33.64  &  1.79 & 0.160  \\  
16202.9956  &  -35.62  &  1.88 & 0.159  \\  
16210.0096  &  -44.80  &  1.66 & 0.159  \\  
16319.6985  &  -83.61  &  1.56 & 0.160  \\  
16327.7120  &  -95.80  &  1.54 & 0.160  \\  
16343.7092  &  -90.62  &  1.66 & 0.155  \\  
16487.1313  &  -122.30  &  1.60 & 0.161  \\  
16508.1360  &  -138.37  &  1.62 & 0.157  \\  
16530.0644  &  -149.89  &  1.58 & 0.157  \\  
16588.9923  &  -138.22  &  2.00 & 0.159  \\  
\enddata
\end{deluxetable}
\capstarttrue

\section{HD~75784} 
\label{sec:hd757}

\subsection{Stellar Characteristics}
\label{sec:757stel}

The \hipp \ catalog lists a parallax of \parseven \  $\pm$ \paruncseven \ mas for HD~75784 (HIP~43569), which corresponds to a distance of \distseven \ $\pm$ \distuncseven \ pc. Spectral synthesis modeling with \sme \ yields \teff \ = \teffseven \ $\pm$ \teffuncseven \ K, \fe = \fehseven \ $\pm$ \fehuncseven \ dex and \logg \ = \loggseven \ $\pm$ \logguncseven. Iteration with the \yy \ isochrones yields a stellar mass of \mstar = \mstseven \ $\pm$ \mstuncseven \ \msun, a stellar luminosity of \lstar = \lstseven \ $\pm$ \lstuncseven \ \lsun, a stellar radius of \rstar = \rstseven \ $\pm$ \rstuncseven \ \rsun, and an age of \ageseven \ $\pm$ \ageuncseven \ Gyr. The measured \logg \ in combination with \teff \ suggests a spectral type and luminosity class most consistent with a \spectypeseven \ subgiant \citep{1966ARA&A...4..193J, 2008oasp.book.....G}. While most of the individual elemental abundances were normal, it is worth noting the low [Si/Fe] abundance of -0.16 $\pm$ 0.05. This low Si abundance is at odds with the result of \citet{2011ApJ...738...97B}, where they found planet hosts to be enhanced in Si relative to Fe. The stellar parameters for HD~75784 are summarized in \autoref{tab:stellar}.

\begin{figure}
\epsfig{file=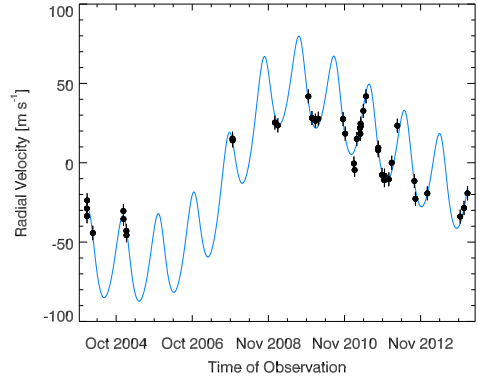,width=1.\linewidth}
\caption{Radial velocity measurements and associated uncertainties for HD~75784 are shown in black with the double Keplerian
model for the two planets superimposed with a solid blue line. }
\label{figHD75784both} \end{figure}

\subsection{Doppler Observations \& Orbital Solution} 
\label{sec:757orb}

Keck HIRES observations of HD~75784 date back to January of 2004, giving a time baseline of ten years for this star. The dominant signal, with an orbital period of \orbperbsvs \ days and eccentricity of \orbebsvs, is significantly longer than our time baseline leading to a poorly-constrained solution using both frequentist and Bayesian approaches. We continue to monitor HD~75784 to refine the orbital solution for this long-period planet; however, with the current set of RV measurements we are able place tight constraints on an additional companion orbiting HD~75784. 

\begin{figure}[b]
\epsfig{file=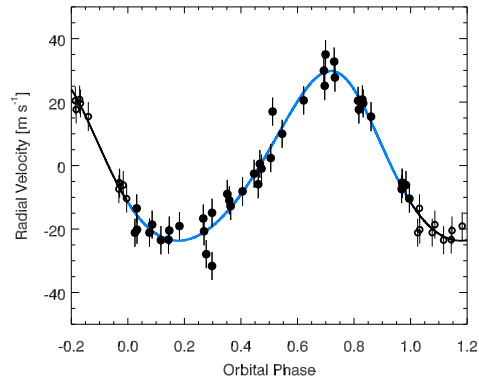,width=1.\linewidth}
\caption{Phased residual RV measurements for HD~75784 using the \orbpercsvs \ orbital period for the inner planet (black) after removing the theoretical model for the outer planet. The blue line shows the best fit Keplerian model for
the innner planet. This shows excellent phase coverage for this solution. }
\label{figHD75784innerphased} \end{figure}

\begin{figure}
\epsfig{file=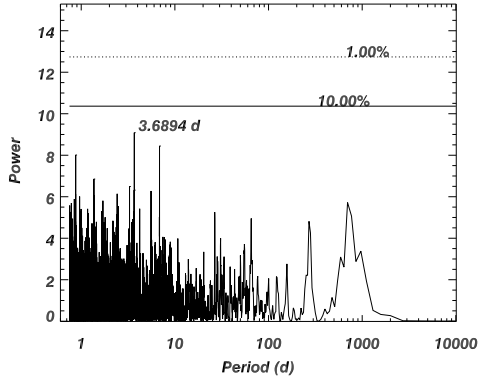,width=1.\linewidth}
\caption{Periodogram of the residuals for HD~75784 after removing the 2 planet Keplerian model. The false alarm probability levels of 10 \% (solid) and 1\% (dotted) are superimposed showing there is no strong periodic signal remaining. This agrees well with our final goodness of fit, showing we did not over-estimate our uncertainty.}
\label{figHD75784pergfap} \end{figure}

Fitting the RV measurements with \kfme \ resulted in an orbital period of \orbpercsvs \ $\pm$ \orbperunccsvs \ days and velocity semi-amplitude of \orbkcsvs \ $\pm$ \orbkunccsvs \ \ms \ for the well-constrained inner planet. Adding the \citet{2010ApJ...725..875I} estimated jitter of 4.3 \ms \ in quadrature to the internal measurement uncertainty resulted in a goodness of fit measurement of 1.25, indicating appropriate estimates for both the jitter and the internal uncertainty. Adopting a stellar mass of \mstseven \ \msun, we derive a semi-major axis of \orbaplcsvs \ $\pm$ \orbaplunccsvs \ AU and a minimum mass of \orbmcsvs \ $\pm$ \orbmunccsvs \ \mjup \ for the inner planet. \autoref{figHD75784both} shows the radial velocity measurements of HD~75784 with the double planet model superimposed in blue. \autoref{figHD75784innerphased} shows the phased model for the inner planet after subtracting the Keplerian model for the outer planet. The Keplerian model for the inner planet was then superimposed in blue, and it can be seen that there is excellent phase coverage. After fitting for both the inner planet and poorly-constrained outer planet, there were no significant signals in the residuals, which can be seen in \autoref{figHD75784pergfap}. The full orbital solution is summarized in \autoref{tab:orb757}. 

\capstartfalse
\begin{deluxetable}{lrr} 
\tablecaption{Orbital Parameters
for the HD~75784 System \label{tab:orb757}}
\tablewidth{0pt} \tablehead{ \colhead{Parameter}  & \colhead{HD 75784~b}
& \colhead{Outer Companion}
} 
\startdata 
P(d) & \orbpercsvs \ $\pm$ \orbperunccsvs & \orbperbsvs \ $\pm$ \orbperuncbsvs    \\ 
T$_{P}$(JD) & \orbtpcsvs \ $\pm$ \orbtpunccsvs & \orbtpbsvs \ $\pm$ \orbtpuncbsvs \\ 
e & \orbecsvs \ $\pm$ \orbeunccsvs & \orbebsvs \ $\pm$ \orbeuncbsvs \\ 
$\omega$ & \orbomcsvs \ $\pm$ \orbomunccsvs & \orbombsvs \ $\pm$ \orbomuncbsvs \\ 
K  (m s$^{-1}$) & \orbkcsvs \ $\pm$ \orbkunccsvs & \orbkbsvs \ $\pm$ \orbkuncbsvs \\ 
a (au) & \orbaplcsvs \ $\pm$ \orbaplunccsvs & \orbaplbsvs \ $\pm$ \orbapluncbsvs \\ 
M $\sin{i}$ (\mjup) & \orbmcsvs \ $\pm$ \orbmunccsvs & \orbmbsvs \ $\pm$ \orbmuncbsvs \\
N$_{\textrm{obs}}$   &   \nobssvs \\
Jitter (m s$^{-1}$)      & \jittersvs \\ 
\rms~(m s$^{-1}$) & \rmssvs \\
$\chi_{\nu}^{2}$ & \chinusvs \\ 
\enddata 
\end{deluxetable}   
\capstarttrue 

\begin{figure}
\epsfig{file=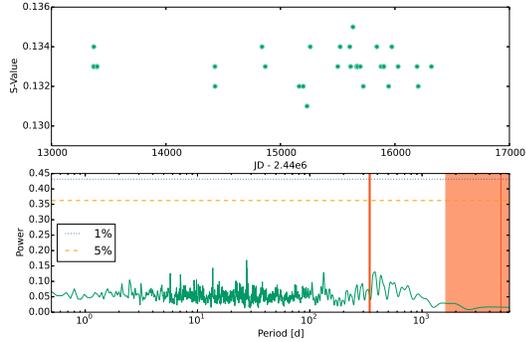,width=1.\linewidth}
\caption{Same as \autoref{figHD5319svals}, but for HD~75784 observations. This shows that there is no indication that the observed signals are related to magnetic activity.}
\label{figHD75784svals} \end{figure}

Similar to the HD~5319 and HD~11506 S-value analysis, we carried out a GLS periodogram analysis of the S-values for the HD~75784 observations. The S-value time series and periodogram are shown in \autoref{figHD75784svals}, where it can be seen that there is power at neither the 342-day nor the 5040-day signals. We also searched for a correlation between the RV measurements and the S-values, resulting in a Pearson correlation coefficient of -0.06 with a p-value of 0.73, indicating there is no correlation between the two parameters.

\begin{deluxetable}{rrrr}[b]
\tablecaption{RV Measurements of HD~75784 \label{tab:rvs75784}}
\tablewidth{0pt}
\tablehead{\colhead{JD}  & \colhead{RV}  
& \colhead{$\sigma_{RV}$} & \colhead{$S_{HK}$}   \\
\colhead{-2440000}  & \colhead{(\ms)}   & \colhead{(\ms)}  } 
\startdata
13014.9237  &  -34.09  &  1.44 &   \\  
13015.9202  &  -38.90  &  1.41 &   \\  
13016.9263  &  -29.04  &  1.44 &   \\  
13071.8866  &  -49.56  &  1.59 &   \\  
13369.0450  &  -35.73  &  1.01 & 0.133  \\  
13369.9123  &  -40.75  &  1.06 & 0.134  \\  
13397.9009  &  -48.22  &  0.82 & 0.133  \\  
13398.8612  &  -51.00  &  0.92 & 0.133  \\  
14428.0495  &  9.88  &  0.90 & 0.133  \\  
14429.0012  &  8.81  &  0.91 & 0.132  \\  
14839.0403  &  20.02  &  1.11 & 0.134  \\  
14867.8771  &  18.26  &  1.16 & 0.133  \\  
15164.0797  &  36.49  &  1.04 & 0.132  \\  
15198.9817  &  22.88  &  0.98 & 0.132  \\  
15231.9936  &  21.20  &  1.00 & 0.131  \\  
15260.7745  &  22.45  &  0.99 & 0.134  \\  
15501.0661  &  22.23  &  1.05 & 0.133  \\  
15522.0041  &  13.04  &  1.01 & 0.134  \\  
15606.0111  &  -5.78  &  1.02 & 0.134  \\  
15613.0127  &  -9.88  &  1.01 & 0.133  \\  
15633.8478  &  9.72  &  0.98 & 0.135  \\  
15663.8356  &  16.54  &  0.93 & 0.133  \\  
15668.8324  &  12.99  &  1.03 & 0.133  \\  
15670.7782  &  19.29  &  0.91 & 0.133  \\  
15672.8363  &  17.69  &  0.95 & 0.133  \\  
15697.7436  &  27.37  &  0.99 & 0.133  \\  
15723.7403  &  36.56  &  1.07 & 0.132  \\  
15842.1218  &  2.62  &  1.06 & 0.134  \\  
15843.0915  &  4.20  &  1.13 & 0.100  \\  
15879.0679  &  -12.94  &  0.95 & 0.133  \\  
15902.0401  &  -16.28  &  0.90 & 0.133  \\  
15902.9937  &  -13.46  &  1.03 & 0.133  \\  
15945.0258  &  -15.73  &  1.08 & 0.132  \\  
15973.0361  &  -5.31  &  1.08 & 0.134  \\  
16027.8576  &  18.07  &  0.98 & 0.133  \\  
16193.1420  &  -16.85  &  0.94 & 0.133  \\  
16203.1417  &  -28.03  &  0.99 & 0.132  \\  
16318.8825  &  -24.60  &  0.88 & 0.133  \\  
16638.0095  &  -39.22  &  1.04 &   \\  
16674.8593  &  -33.82  &  1.33 &   \\  
16708.9338  &  -24.55  &  1.10 &   \\  
\enddata
\end{deluxetable}

\section{HD~10442}
 \label{sec:hd104}

\subsection{Stellar Characteristics}
\label{sec:104stel}

\begin{figure}[t]
\epsfig{file=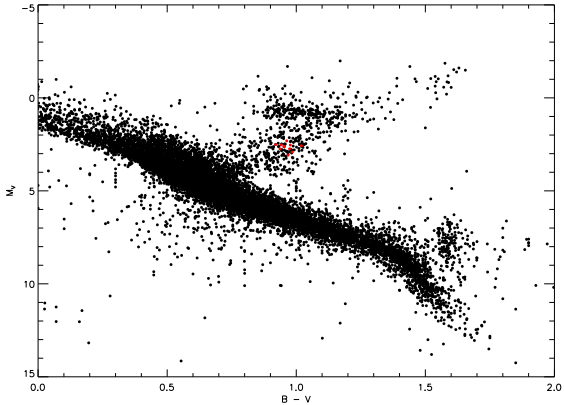,width=1. \linewidth,clip=}
\caption{HR diagram of the stars within 50 pc from the \hipp \ Catalog. Superimposed in red are the stars used to estimate the mass of HD~10442.} 
\label{fig10442hrspocs}
\end{figure}

As stated in \S \ref{sec:n2k}, the N2K sample was selected from the \hipp \ Catalog. Since HD~10442 (TYC~32-383-1) is not a member of the \hipp \ catalog, this star is likely one of a few metal-rich stars that were added to the target list as part of an undergraduate research project. Without knowing the distance to HD~10442, we could not iterate between the \yy \ isochrones and \sme \ as described in \S \ref{sec:anal}, and therefore do not have values for the stellar mass, luminosity, age, or stellar radius as we do for the other 3 stars presented in this work. The stellar characteristics calculated using the non-iterative form of \sme \ for HD~10442 are \logg \ = \loggten \ $\pm$ \logguncten, \teff \ = \tefften \ $\pm$ \teffuncten \ K, and \fe \ = \fehten \ $\pm$ \fehuncten.

\begin{figure}[t]
\epsfig{file=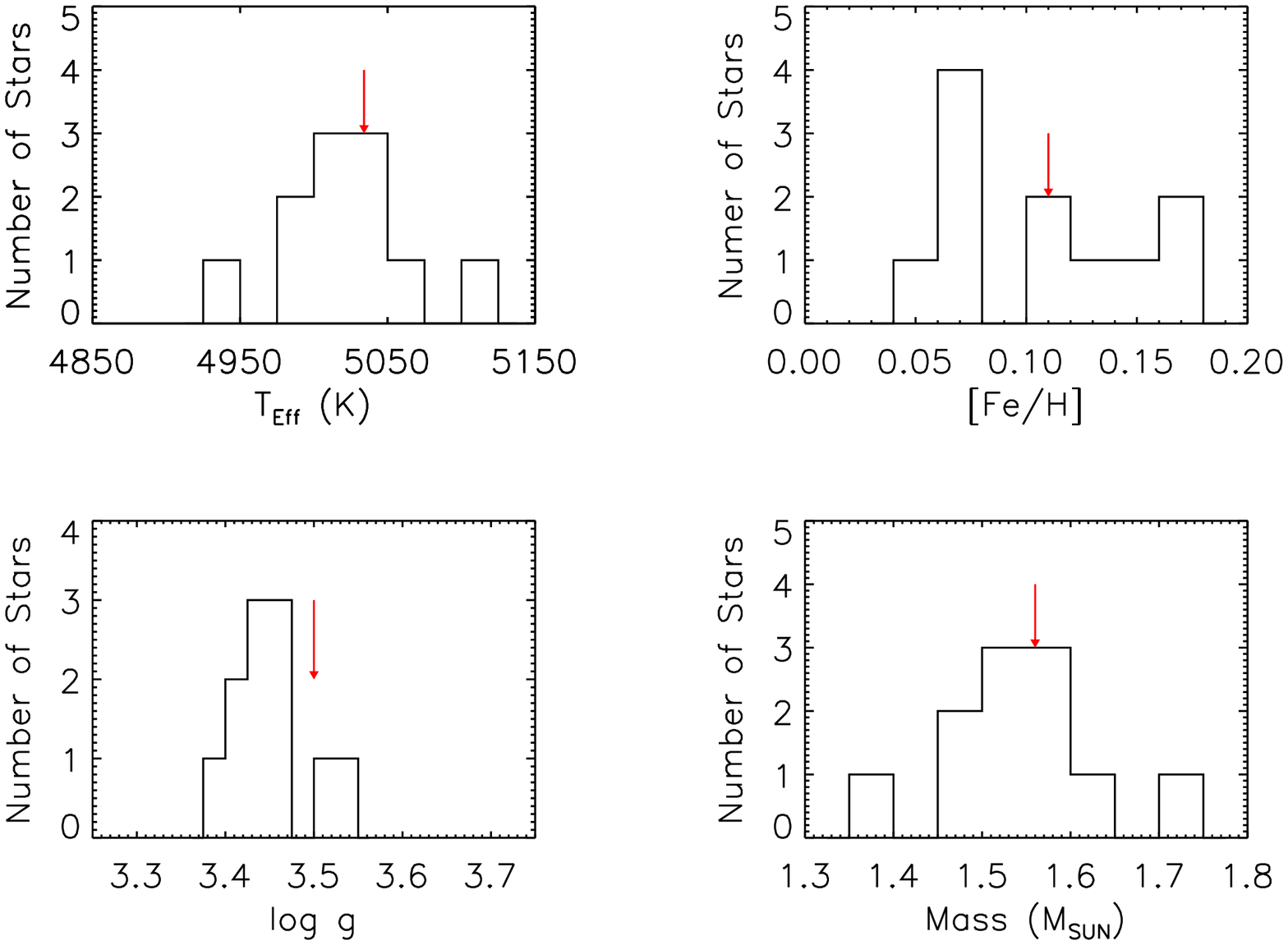,width=0.95\linewidth,clip=}
\caption{\teff, \fe, \logg, and mass distributions for the 11 stars within 2-$\sigma$ of the \teff, , \fe, and \logg \ of HD~10442. The red arrows show the values for HD~10442.} 
\label{fig10442spocsdist}
\end{figure}

\begin{figure}[t]
\epsfig{file=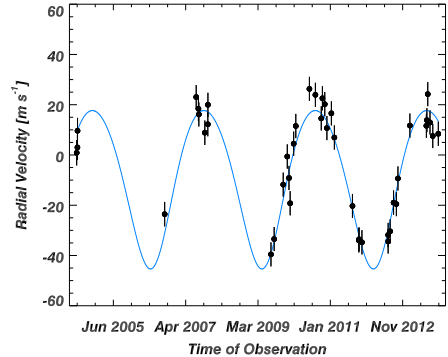,width=0.95\linewidth,clip=}
\caption{Radial velocity measurements for HD~10442 with associated errors (black) and the best-fit Keplerian model superimposed in blue.} 
\label{fig10442rvs} 
\end{figure}

To estimate the stellar mass of HD~10442, we searched our SME analysis of stars that were within 2-$\sigma$ of the \sme \ derived \logg, \fe, and \teff \ of HD~10442. The resulting 11 stars satisfying the 2-$\sigma$ criteria are shown in red in \autoref{fig10442hrspocs} amongst all stars in the \hipp \ Catalog that are within 50 pc of the Sun. The median stellar mass of the 11 star sample  (\mstten \ \msun) and standard deviation  (\mstuncten) were adopted for the mass and associated uncertainty of HD~10442 when calculating the orbital parameters of HD~10442~b. \autoref{fig10442spocsdist} shows the \teff, \fe, \logg, and stellar masses of these 11 stars, and the red arrows show the values for HD~10442. 

The Tycho B$_{T}$ and V$_{T}$ magnitudes for HD~10442 are 9.06 and 7.94, respectively. Converting to Johnson magnitudes gives V~=~7.84 and B~-~V~=~0.93. SIMBAD lists HD~10442 as a G5 star of unknown luminosity class. Based on the stellar parameters derived using \sme, the B and V values from Tycho, and the position on the HR diagram, this star most closely resembles a \spectypeten \ subgiant. We therefore adopt a jitter value of 4.7 \ms \ from \citet{2010ApJ...725..875I}. The stellar parameters for HD~10442 are summarized in \autoref{tab:stellar}.

\begin{figure}[b]
\epsfig{file=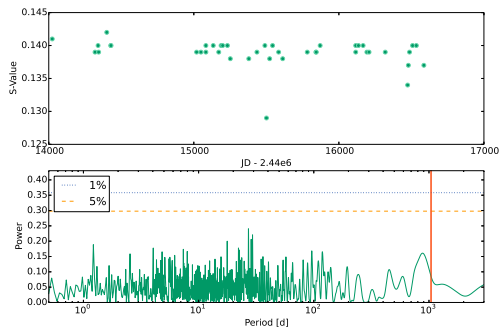,width=0.95\linewidth,clip=}
\caption{Same as \autoref{figHD5319svals}, but for S-values of HD~10442 observations. This shows that there is no significant magnetic activity signal.} 
\label{figHD10442svals} 
\end{figure}

\subsection{Doppler Observations \& Orbital Solution} 
\label{sec:104orb}
HD~10442 was first observed with the HIRES spectrometer at Keck Observatory in July of 2004. Although it was clear after the first few observations that this star did not harbor a hot Jupiter, the velocities showed a significant linear trend. HD~10442 was therefore kept on the active observing program. Now, with a time baseline of Doppler measurements spanning more than 10 years, the planetary nature of this signal has been confirmed. 

The orbital solution that best fits the Doppler measurements has an orbital period of \orbperbozu \ $\pm$ \orbperuncbozu \ days, an eccentricity of \orbebozu \ $\pm$ \orbeuncbozu, and semi-amplitude of \orbkbozu \ $\pm$ \orbkuncbozu \ \ms. Using these parameters and assuming a stellar mass of \mstten \ \msun, we calculate a minimum mass of \orbmbozu \ $\pm$ \orbmuncbozu \ \mjup \ and a semi-major axis of \orbaplbozu \ $\pm$ \orbapluncbozu \ AU for the planetary companion. \autoref{fig10442rvs} shows the full set of radial velocity measurements and the best-fit single-planet orbital solution is superimposed with a solid blue line. Both the periodogram and Keplerian FAP analyses give an FAP of $<$ 0.1\% for HD~10442~b. 

As with the other systems, periodogram analysis of the S-values for the HD~10442 observations (shown in \autoref{figHD10442svals}) reveals no significant power due to magnetic activity. A search for a linear correlation between the RV measurements results in a Pearson correlation coefficient of -0.28 with a corresponding p-value of 0.08 -- again showing no significant correlation between the two parameters -- supporting the planetary interpretation.

\begin{figure}
\epsfig{file=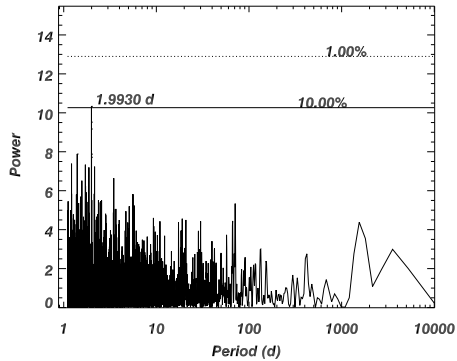,width=0.95\linewidth,clip=}
\caption{Periodogram for the residuals of HD~10442 after subtracting the 
best-fit model, Superimposed are FAP levels of 10 \% (solid), 1\% (dotted) 
and 0.1\% (dashed). This shows a signal with a period of $\sim$ 2 days, but with a high FAP
indicating it is most likely due to a window function in the time series data.}
\label{figHD10442residpergfap} 
\end{figure}

\begin{figure}[b]
\epsfig{file=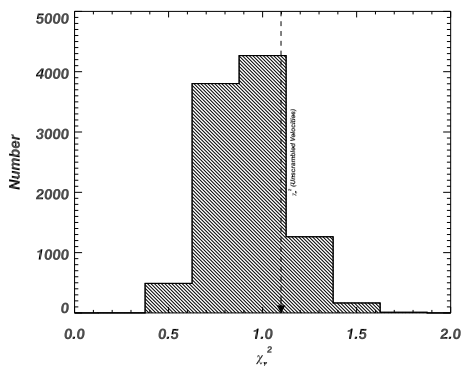,width=0.95\linewidth,clip=}
\caption{Best-fit \chinu \ distribution after performing a bootstrap Monte
Carlo scrambling of the residual velocities to the 2 planet fit for HD~10442,
then refitting with a Keplerian model. $>$80\% of the 10,000 realizations resulted in
a lower \chinu~than the unscrambled velocities, reiterating the periodogram FAP analysis
result that the 2-day signal is a window function in the data.} 
\label{fig10442ckepfap}
\end{figure}

Analysis of the residuals to this single-planet solution shows a spike of power in the residuals corresponding to a period of $\sim$~2 days. However, Figures \ref{figHD10442residpergfap} and \ref{fig10442ckepfap} show that the periodogram and Keplerian FAP tests both give a high FAP for this period. Furthermore, the \chinu \ did not improve when including the 2 day signal in the orbital solution. This leads us to conclude that this 2 day signal is not due to the presence of an additional companion, but rather a window function in our radial velocity set. The full orbital solution for HD~10442~b is summarized in \autoref{tab:orb104}.

\begin{deluxetable}{lr} 
\tablecaption{Orbital Parameters for the HD~10442 System \label{tab:orb104}}
\tablewidth{0pt} 
\tablehead{ \colhead{Parameter}  & \colhead{HD 10442~b}
} 
\startdata 
P(d) & \orbperbozu \ $\pm$ \orbperuncbozu    \\ 
T$_{P}$(JD) &
\orbtpbozu \ $\pm$ \orbtpuncbozu \\ 
e & \orbebozu \ $\pm$ \orbeuncbozu\\
$\omega$ & \orbombozu \ $\pm$ \orbomuncbozu \\ 
K  (m s$^{-1}$) & \orbkbozu \ $\pm$ \orbkuncbozu \\ 
a (au) & \orbaplbozu \ $\pm$ \orbapluncbozu \\ 
M$\sin{i}$  (\mjup) & \orbmbozu \ $\pm$ \orbmuncbozu \\ 
N$_{\textrm{obs}}$   &   \nobsozu \\ 
Jitter (m s$^{-1}$)      & \jitterozu \\ 
\rms~(m s$^{-1}$) & \rmsozu \\ 
$\chi_{\nu}^{2}$ & \chinuozu \\ 
\enddata 
\end{deluxetable}


\section{Summary \& Discussion} 
\label{sec:disc}

Here we have presented 4 newly-discovered exoplanets (HD~5319~c, HD~11506~c, HD~75784~b and HD~10442~b), refined orbital parameters for 2 previously published planets (HD~5319~b and HD~11506~b), and have shown indications that two more companions may exist (orbiting HD~11506, HD~75784) but need additional observations to constrain their orbital parameters. Two of these stars (HD~5319 and HD~11506) were already known to harbor single gas giant planets. These two systems have therefore transitioned from the ensemble of known single-planet systems to multi-planet systems. The detection of additional planets orbiting these stars supports the result of \citet{2009ApJ...693.1084W}, where they found that the most probable multi-planet systems are systems where single planets have already been detected.

HD~5319 is a remarkable system due to its unknown formation mechanism. Through hydrodynamical simulations \citet{2012MNRAS.426..187R} have shown that massive planets cannot form in situ in the 4:3 resonance. Instead, these planets must have undergone migration to get to their current positions. However, \citet{2012MNRAS.426..187R} went on to show that convergent migration also fails to create high mass planetary systems in 4:3 resonances because unphysical migration rates are needed to overcome the more widely separated first order resonances. In the same work \citet{2012MNRAS.426..187R} also showed that it is unlikely that the number of observed gas giant systems in the 4:3 resonance could have been created through planet-planet scattering. While they suggest two unexplored possibilities for the formation of high mass planets in 4:3 MMRs, the formation mechanism of the HD~5319 system is currently an open problem.

HD~11506 has also been promoted to multi-planet status. The outer planet orbiting HD~11506 was first announced by \citet{2007ApJ...669.1336F}. \citet{2007ApJ...669.1336F} also commented that several peaks existed in the periodogram of the residuals, including a peak in the power at 170 days; however, they stated more data were needed to evaluate the second-signal. Reanalyzing the RV measurements from that work, \citet{2009A&A...496L..13T} claimed the period of the second planet was 170.5$^{+3.3}_{-6.2}$ with 99\% confidence. With the additional 87 observations presented in this work, we find an orbital period for the second planet of \orbpercoov \ $\pm$ \orbperunccoov \ days, which is significantly different than 170-day signal that was starting to become apparent in the previously-published data set by \citet{2007ApJ...669.1336F}. The best-fit solution now has two planets with well-constrained orbital parameters and a distant third companion approximated as a linear contribution to the radial velocity measurements.

\begin{deluxetable}{rrrr}[t]
\tablecaption{RV Measurements of HD~10442 \label{tab:rvs10442}}
\tablewidth{0pt}
\tablehead{\colhead{JD}  & \colhead{RV}  
& \colhead{$\sigma_{RV}$} & \colhead{$S_{HK}$}   \\
\colhead{-2440000}  & \colhead{(\ms)}   & \colhead{(\ms)}  } 
\startdata
13200.0655  &  0.86  &  1.83 &   \\  
13207.0955  &  3.01  &  1.75 &   \\  
13208.0761  &  9.64  &  2.02 &   \\  
14024.0035  &  -23.56  &  1.24 & 0.141  \\  
14319.0890  &  23.06  &  0.78 & 0.139  \\  
14339.9948  &  18.39  &  0.99 & 0.140  \\  
14343.9694  &  16.15  &  0.98 & 0.139  \\  
14399.9071  &  8.86  &  1.29 & 0.142  \\  
14427.9188  &  12.22  &  0.99 & 0.140  \\  
14428.8102  &  19.97  &  0.94 & 0.140  \\  
15019.1203  &  -39.55  &  0.99 & 0.139  \\  
15049.1033  &  -33.44  &  0.94 & 0.139  \\  
15133.9656  &  -11.80  &  0.99 & 0.140  \\  
15171.8824  &  -0.61  &  1.12 & 0.139  \\  
15187.8836  &  -9.16  &  1.04 & 0.140  \\  
15198.8405  &  -19.21  &  1.04 & 0.140  \\  
15231.7423  &  4.49  &  1.07 & 0.140  \\  
15251.7110  &  11.53  &  1.05 & 0.138  \\  
15379.1166  &  26.31  &  0.89 & 0.138  \\  
15435.1172  &  23.94  &  0.99 & 0.139  \\  
15489.9581  &  14.58  &  1.08 & 0.140  \\  
15500.8787  &  22.56  &  1.01 & 0.129  \\  
15522.9023  &  20.21  &  1.09 & 0.138  \\  
15542.9554  &  10.71  &  0.96 & 0.140  \\  
15584.7897  &  16.57  &  1.01 & 0.139  \\  
15612.7098  &  6.95  &  1.08 & 0.138  \\  
15782.1119  &  -20.30  &  0.92 & 0.139  \\  
15841.9779  &  -33.99  &  0.95 & 0.139  \\  
15843.9846  &  -33.55  &  0.94 & 0.139  \\  
15870.9958  &  -34.75  &  1.13 & 0.140  \\  
16116.1286  &  -34.38  &  1.20 & 0.140  \\  
16116.1295  &  -31.85  &  1.16 & 0.139  \\  
16135.1267  &  -30.37  &  0.98 & 0.140  \\  
16167.1030  &  -18.96  &  1.12 & 0.140  \\  
16193.1258  &  -19.54  &  1.00 & 0.139  \\  
16207.9660  &  -9.35  &  0.95 & 0.139  \\  
16319.7270  &  11.70  &  0.96 & 0.139  \\  
16474.1247  &  11.62  &  1.00 & 0.134  \\  
16479.1294  &  13.86  &  0.89 & 0.137  \\  
16487.1288  &  24.21  &  0.95 & 0.139  \\  
16508.1309  &  12.95  &  0.91 & 0.140  \\  
16534.0526  &  7.54  &  0.97 & 0.140  \\  
16585.9211  &  8.44  &  1.13 & 0.137  \\  
\enddata
\end{deluxetable}

An interesting characteristic of the HD~11506 system is the linear correlation between the residual velocities after subtracting the 2-planet + linear trend model and the S-values. Subtracting this magnetic signal from the velocities and refitting the system had no significant effect on the best-fit orbital parameters, but it did lower the rms by 1.0 \ms. This emphasizes the need for sophisticated methods to handle stellar activity when searching for low mass planets.

A planet has also been discovered orbiting HD~75784, which was not known to host any planets prior to this work. To properly model the radial velocity measurements of HD~75784, a second (longer period) Keplerian signal needed to be included. However, the time baseline of our radial velocity measurements is shorter than the orbital period for the outer planet leading to a poorly-constrained orbital solution using both frequentist and Bayesian approaches. Although the orbital solution for the outer planet is poorly constrained, the solution for the inner planet is well-known and warrants publication at this time. HD~75784 will remain an active target to constrain the orbital parameters for the outer planet. The current best-fit solution for the outer planet has a semi-major axis of $\sim$6.5 AU, making it one of the most widely-separated planets discovered with the radial velocity technique. An interesting characteristic of HD~75784 is that it has an abnormally low [Si/Fe] of -0.16 $\pm$ 0.05. \citet{2011ApJ...738...97B} found that gas giants are preferentially detected orbiting stars that are enhanced in silicon relative to iron, making the HD~75784 system a curious outlier to their observations. 

Lastly, the we announced a single gas giant orbiting HD~10442. Unlike the rest of the stars discussed in this work, HD~10442 does not have a \hipp \ parallax measurement and we could therefore not use the \yy~isochrones to determine its mass. To calculate the mass of HD~10442~b, we instead used the median mass of stars from a modified SPOCS catalog that are similar to HD~10442 in \feh, \teff \ and \logg. 

\begin{figure}[b]
\epsfig{file=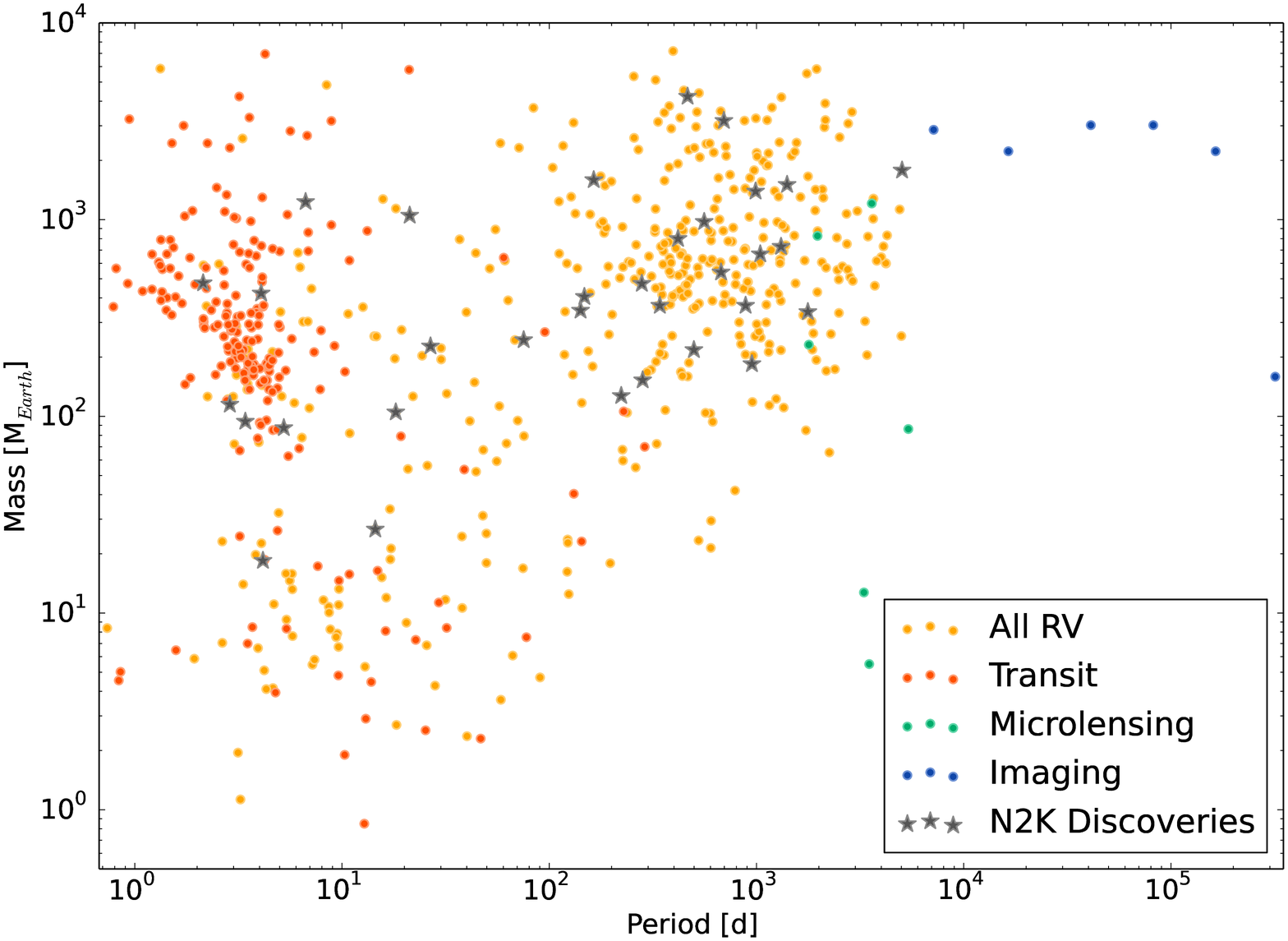,width=1.0\linewidth}
\caption{All confirmed planets listed on exoplanets.org. Planets discovered using the transit, RV, microlensing, and direct imaging techniques are shown in red, orange, green, and blue circles, respectively. Superimposed in gray stars are the planets discovered through the N2K Consortium.} 
\label{figmassper} 
\end{figure}

These discoveries bring the total number of planets detected through the N2K Consortium to 32. The original goal of the N2K was to detect short-period gas giants, which have high transit probabilities. It has since evolved into a campaign to detect long-period planets. The full ensemble of planets discovered through the N2K Program is shown in \autoref{figmassper} (gray star symbols) amidst all known exoplanets listed on \href{http://exoplanets.org}{exoplanets.org}. This shows the wide-range of mass-period parameter space covered by planets discovered through this program. Included as a member of the N2K detections in \autoref{figmassper} is the candidate HD~75784~c. While this outer companion orbiting HD~75784 is still poorly-constrained, additional observations of this target, and many others that are from the original pool of stars observed as part of N2K, will help build a large population of widely separated planets discovered with the radial velocity method. Building a large population of such widely separated systems will be useful for the future comparison of the occurrence rate of planets discovered using the direct imaging method and radial velocity method, and to refine population synthesis models to improve our understanding of planet migration. 

\acknowledgements We thank the referee for useful comments and suggestions. We are grateful to Geoff Marcy for his longstanding contributions to this program and for helping with the observing and data reduction. We also thank Camille Avestruz, Ana Bonaca, Kristina Douglas, Aurelia Giguere, Haven Giguere, Nicole Larsen, Jack Moriarty, and Adele Plunkett for useful discussions. We gratefully acknowledge the dedication and support of the Keck Observatory staff, in particular Grant Hill and Scott Dahm for their support with HIRES and Greg Wirth, Bob Kibrick, Craig Henry and Andrew Stemmer for supporting remote observing. We thank the NASA and Yale Telescope assignment committees for generous allocations of telescope time. This work was supported by NASA Headquarters under the NASA Earth and Space Science Fellowship Program - Grant NNX13AM15H. Fischer acknowledges support from NASA grant NNX08AF42G and NASA Keck PI data analysis funds. This research has made use of the Extrasolar Planets Encyclopedia, available at \href{http://exoplanet.eu}{exoplanet.eu}, and the Exoplanet Orbit Database at \href{http://exoplanets.org}{exoplanets.org}. This research has also made use of the \href{http://simbad.u-strasbg.fr/simbad/sim-fid}{SIMBAD} database, operated at CDS, Strasbourg, France. Data presented herein were obtained at the W. M. Keck Observatory from telescope time allocated to the National Aeronautics and Space Administration through the agency's scientific partnership with the California Institute of Technology and the University of
California.  The Observatory was made possible by the generous financial support of the W. M. Keck Foundation.  The authors extend thanks to those of native Hawaiian ancestry on whose sacred mountain of Mauna Kea we are privileged to be guests.  Without their generous hospitality, the Keck observations presented herein would not have been possible.

\bibliography{ms.bib}
\end{document}